\documentclass[10pt,final,3p,times,fleqn,lefttitle]{elsarticle}

\usepackage{fancyhdr}
\usepackage{amsmath,bm}
\usepackage{amssymb}
\usepackage{graphicx}
\usepackage{caption}
\usepackage{subcaption}
\usepackage{float}
\usepackage[colorlinks]{hyperref}
\begin{document}
	\begin{frontmatter}	
		\title{Contact radius and curvature corrections to the nonlocal contact formulation accounting for multi-particle interactions in elastic confined granular systems}
		\author[1]{Ankit Agarwal}
		\ead{agarwa80@purdue.edu}
		\author[1,2]{Marcial Gonzalez\corref{cor2}}
		\ead{marcial-gonzalez@purdue.edu}
		\cortext[cor2]{Corresponding author. Tel.: +1 765 494 0904; fax: +1 765 496 7537}
		\ead[url]{www.marcialgonzalez.net}
		
		\address[1]{School of Mechanical Engineering, Purdue University, West Lafayette, IN 47907, USA}
		\address[2]{Ray W. Herrick Laboratories, Purdue University, West Lafayette, IN 47907, USA}	
		\begin{abstract}
			We present contact radius and curvature corrections to the nonlocal contact formulation that account for multi-particle interactions in elastic confined granular systems. The nonlocal contact formulation removes the classical assumption of independent contacts by taking into account the interplay of deformations due to multiple contact forces acting on a single particle. The contact radius correction considers the components of these deformations that contribute to the inter-particle contact area. The curvature correction improves the description of contacting surface profiles by including higher order terms in their Taylor series expansions. To validate the corrected formulation, we restrict attention to rubber spheres under different loading conditions, in the absence of gravitational forces, adhesion or friction. Specifically, we show that the predictions of contact force and radius are in remarkable agreement with finite-element simulations and experimental observations up to levels of deformation at which contact impingement occurs, which was not possible with the original elastic nonlocal contact formulation. Convergence of the curvature corrected formulation is observed at a four-term correction.
		\end{abstract}
		
		
	\end{frontmatter}

	\section{Introduction} \label{sec1}
	The extensive applications of powder compaction, especially in manufacturing processes of critical industries like pharmaceutics, ceramic, energy, food, and metallurgy, make it a subject of intense research in the scientific community. Development of predictive and computationally efficient models that could accurately describe the behavior of granular media during compaction would directly impact optimality in manufacturing, waste reduction, and price and quality of the end product.   
	
	Macroscopic behavior of confined granular systems has been conventionally described by \textit{continuum-based models}, which consider granular media as a continuous system and hence have a minimal emphasis on the behavior at particle scale. Many of these models were originally developed for analyzing the behavior of geological materials, such as \cite{DruckerPrager-1952}, Cam-Clay plasticity and Cap plasticity models. More recently, the Drucker-Prager Cap (DPC) plasticity model \citep*{Dimaggio-1971}, where a cap yield surface is added to the \citeauthor{DruckerPrager-1952} model to allow for material hardening and dilatancy control during inelastic deformation, has been used for analysis of metal, ceramic and pharmaceutical powder compaction. Although requiring an elaborate mechanical testing procedure for calibration of model parameters \citep*{Cunningham-2004}, the DPC model is widely used due to its adaptability to finite element method \citep*{Sinka-2004}. However, the accuracy of model response, especially during the decompression (unloading) phase, relies heavily on the design of calibration experiments \citep*{Sinha-2010} and proper elastic constitutive modeling \citep*{Han-2008}. In order to incorporate microstructural properties of the granular system into its global behavior, \textit{discrete models} have been proposed, where contact behavior of individual particles is taken into account. Numerical methods in this category, such as dynamic discrete element methods \citep*{Cundall-1979, zhu2008discrete} and quasi-static particle mechanics approaches \citep*{Gonzalez-2016, yohannes2016evolution, Gonzalez-2018}, are used in combination with a suitable contact formulation to predict the macroscopic behavior of compacted granular systems and, thus, predictability is heavily dependent on the contact law involved. The \cite{Hertz-1882} contact law for linear-elastic materials and similarity solution by \citet*{Storakers-1997} for viscous-plastic power law hardening materials are fairly predictable at small deformations and low relative densities of powder compacts. However, due to the occurrence of contact interactions at higher deformations, as pointed out by \cite{Mesarovick-2000} in their study of elasto-plastic spheres, their predictions become increasingly deviant due to the assumption of independent contacts. This was partially overcome by introducing a local relative density parameter in contact laws curve-fitted to finite element simulations of small three-dimensional packings \citep*{Harthong-2009}. Finally, a systematic and mechanistic connection between macroscopic and particle level behaviors, using continuum and discrete models respectively, was recently proposed \citep*{Poorsolhjouy-2018} to capture the anisotropic evolution of die-compacted systems.
	
    Several efforts towards experimental characterization of confined granular systems have also been made to understand the deformation behavior at granular scale and to provide an efficient validation tool for the analytical contact formulations. Of particular interest is the mechanical response of single particles under confined conditions, most commonly studied using uniaxial compression experiments \citep*{Liu-1998,Lu-2001,Shima-1993,Tatara-1991b,Topuz-2009,Zhang-2006}. Recently,  an apparatus has been developed for triaxial testing of single particles \citep*{Jonsson-2015}, providing a more realistic insight into the behavior of individual particles during powder compaction.

	For elastic confined granular systems, relaxing the underlying assumptions of the Hertz contact theory that limit its applicability to small deformations could be the key to achieving predictability at moderate to large deformations. Significant efforts in this direction were made by \cite{Zhupanska-2011}, who relaxed the small-strain Hertz assumption of considering contacting surfaces as elastic half spaces by proposing an analytical solution to the boundary value problem of an elastic sphere subject to contact stresses on a finite region of its surface and supported at its center. The results showed that the Hertz pressure distribution remained accurate for relatively large contact areas. Recently, \cite{Argatov-2017} have explored the concept of far points in Hertz contact problems, emphasizing the limitations of the "local character" of Hertz predictions. Another major contribution in this regard is the nonlocal contact formulation for confined granular systems by \cite{Gonzalez-2012}, which provides an accurate and mechanistic description of the force-deformation behavior at contacts of a linear-elastic spherical particle subject to multiple contact forces, a typical configuration in particulate systems compressed to high relative densities. It follows the work of \cite{Tatara-1989} and relaxes the classical contact mechanics assumption of independent contacts by invoking the principle of superposition to express the deformation at a particular contact as a sum of local (i.e., Hertzian) deformation and nonlocal deformations generated by other contact forces acting on the same particle. The nonlocal contact formulation has recently been employed successfully to study the die-compaction of large frictionless noncohesive granular systems comprising weightless elastic spherical particles \citep{Gonzalez-2016}.
	
	A complete description of the inter-particle contact behavior in confined granular systems includes determination of both contact force and area with respect to particle deformation. While critical macroscopic quantities like compaction pressure and the reaction from die walls are directly related to inter-particle contact forces, the prediction of contact area is needed to estimate strength formation in the compacted solid \citep*{Gonzalez-2018b}. In addition, the evolution of contact area is associated with contact impingement, i.e., with the merger of neighboring contacts. Since the assumption of circular contacts no longer remains valid after contact impingement, the predictions of a contact formulation may not be representative of real contact behavior beyond the occurrence of this phenomenon, making an accurate determination of contact areas ever so important.
	
	In the context of the nonlocal contact formulation \citep*{Gonzalez-2012}, nonlocal mesoscopic deformations are derived from the Boussinesq solution \citep*{Johnson-1985,Timo-1970} of an elastic half-space under a concentrated force. The components of these deformations normal to the contact surface constitute the nonlocal contribution to the contact displacement, for which a closed-form solution has been obtained \citep*{Gonzalez-2012}. However, the derivation of an analytical solution for nonlocal components radial to the contact center, that contribute to the evolution of contact radius, remains an open problem. Therefore, part of the work presented in this paper is concerned with the development of an analytical framework for predicting nonlocal effects in the evolution of inter-particle contact area. The analysis presented is in the spirit of \citeauthor{Tatara-1991a}'s (\citeyear{Tatara-1991a}) work on expanded contact radius during uniaxial compression of rubber spheres.
	
	An important aspect of this nonlocal contact formulation is that it is a direct extension of the classical Hertz contact theory. When nonlocal effects are neglected, the formulation reduces to Hertz theory. Therefore, any correction introduced in the Hertz solution should, in turn, improve the accuracy of the formulation. The second analysis presented in this paper is concerned with the improvement of the formulation using contact pressures of higher accuracy obtained by correcting the description of the profiles of contacting surfaces through higher order terms in the Taylor series expansion of the profile functions. This methodology, termed curvature correction, is similar to the one reported by \cite{Cattaneo-1947} for solids of revolution and \cite{LUO-1958} for general solids. Closed form solutions of contact force and radius in terms of displacement are obtained for two-, three- and four-term curvature corrections. 
	
	Finally, a validation of the two proposed corrections is performed by comparison of the analytical predictions of contact force and radius versus deformation for compression of rubber spheres under various loading configurations with finite element simulations and experimental measurements. 
	
	The paper is organized as follows. The deformation of an elastic sphere in a confined granular system under the action of multiple contact forces is discussed in Section \ref{sec2}. The analytical formulation for predicting nonlocal effects in the evolution of contact radius is then presented in Section \ref{sec3}. In Section \ref{sec4}, a curvature corrected nonlocal contact formulation is discussed. Section \ref{sec5} presents a numerical and experimental validation of contact radius and curvature corrections. Finally, a summary and concluding remarks are presented in Section \ref{sec6}.
	
	\begin{figure}[t]
		\centering
		\includegraphics[keepaspectratio,scale=0.55]{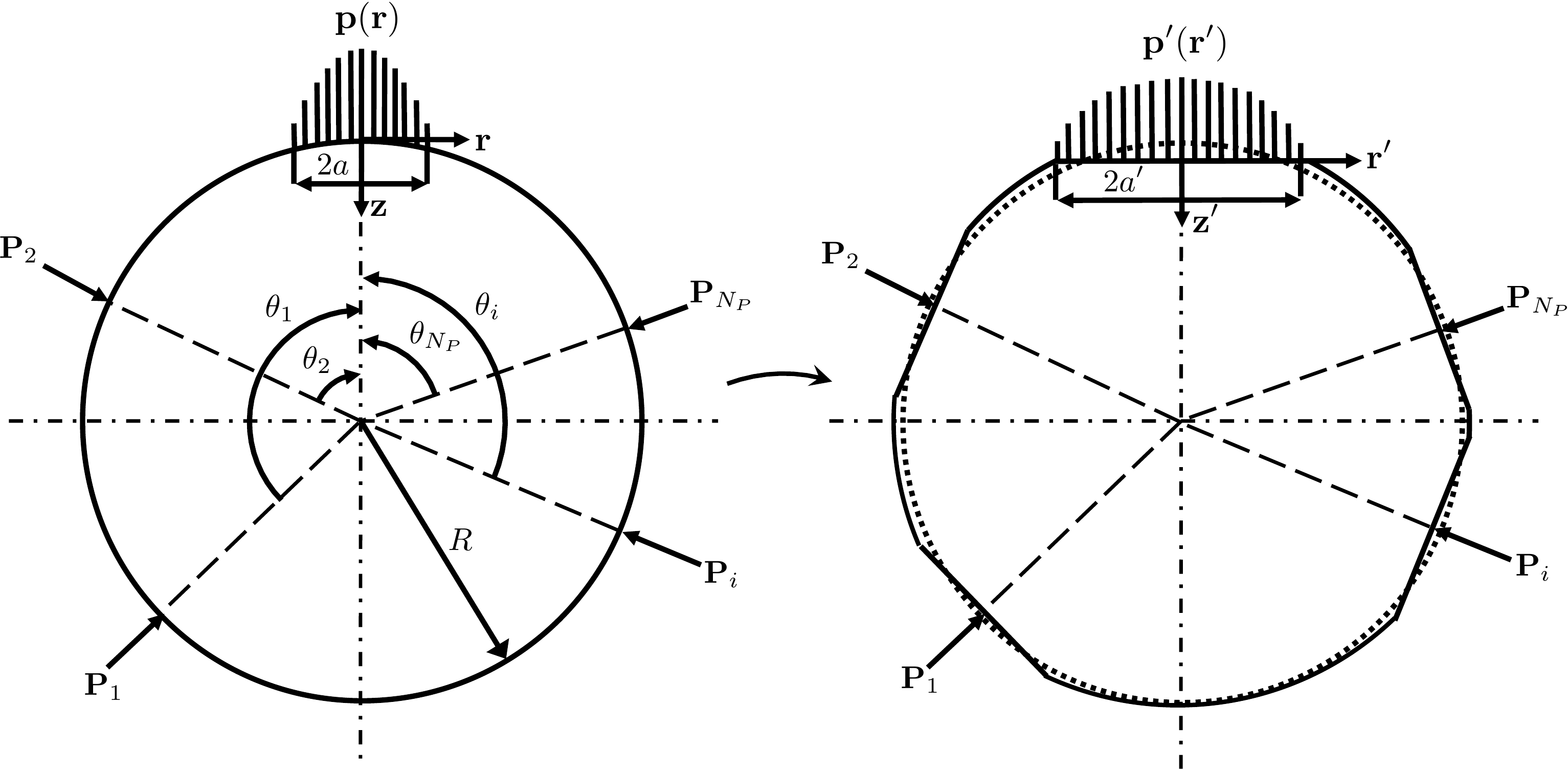}
		\caption{Depiction of deformation of an elastic sphere under the action of a general configuration of contact forces, due to a distributed surface pressure $p$ acting on its surface. The figure on the left denotes the configuration before deformation (reference configuration) while the figure on the right denotes the deformed configuration.}
		\label{fig:deformation}
	\end{figure}	

	\section{Deformation of an elastic sphere in a confined granular system under the action of a distributed surface pressure} \label{sec2}
	
	We consider the deformation of a linear elastic sphere with radius $R$, Young's modulus $E$ and Poisson's ration $\nu$, under the action of a distributed pressure and a general loading configuration of multiple contact forces. By approximating the deformation with an axially symmetric field, a cylindrical coordinate system defined in Figure~\ref{fig:deformation} is adopted. The reference coordinates are denoted by $\mathbf{X}:(r, z)$, while the deformed coordinates are denoted by $\mathbf{x}: (r', z')$. The general loading configuration is represented by $(P_i, \theta_i)$, where $P_i$ are the forces acting on the sphere's surface and $\theta_i$ are their angular distances with respect to $z$ or $z'$ axis. 
	
	A spherical cap of base radius $a$ deforms to a flat circular surface of radius $a'$ under the action of an ellipsoidally distributed pressure $p$ proposed by \cite{Hertz-1882} and given by
	\begin{equation} \label{1}
	p(r) = \bar{p}_m\sqrt{1-\frac{r^2}{a^2}}
	\end{equation}
	where $\bar{p}_m=3P/2\pi a^2$ is the maximum value of the pressure and $P$ is the effective contact force. The displacement of the elastic cap is represented by a deformation mapping $\bm{\varphi(\mathbf{X})}$, given by
	\begin{equation} \label{2}
	\begin{aligned}
	\begin{Bmatrix}
	r' \\
	z'
	\end{Bmatrix} = \bm{\varphi(\mathbf{X})}= 
	\begin{Bmatrix}
	r+u(r)+\sum\limits_{i=1}^{N_P} u_{P_i}(r)\\
	R - \sqrt{R^2 - r^2} + w(r) - \sum\limits_{i=1}^{N_P} w_{P_i}(r)
	\end{Bmatrix} \\
	\end{aligned}
	\end{equation}
	where $z = R - \sqrt{R^2 - r^2}$ is used, thus enabling the deformation mapping of surface points solely in terms of $r$. Quantities $w(r)$ and $u(r)$ are the vertical and radial displacements of the cap's surface points due to local pressure $p(r)$, which can be approximated by means of Boussinesq solution \citep*{Johnson-1985,Timo-1970}, i.e., by
	\begin{equation} \label{3}
	w(r) = \frac{3P(1-\nu^2)}{8a^3E}(2a^2-r^2)
	\end{equation}
	and, according to \cite{Tatara-1991a}, by
	\begin{equation} \label{4}
	u(r) = \frac{(1+\nu)P}{4\pi Er}\left[\frac{r}{\sqrt{2}R}\sqrt{1+\sqrt{1-\left(\frac{r}{R}\right)^2}}-2(1-2\nu)\left(1-\frac{1}{\sqrt{2}}\sqrt{1-\sqrt{1-\left(\frac{r}{R}\right)^2}}\right)\right]
	\end{equation}
	Quantities $u_{P_i}(r)$ and $w_{P_i}(r)$ are the nonlocal contributions to radial and vertical displacements induced by concentrated forces $P_i$ acting on the surface of the sphere. As a valid approximation, \cite{Gonzalez-2012} have represented $w_{P_i}(r)$ by the value of $w_{P_i}$ at the contact center, i.e., by  
	\begin{equation} \label{5}
	w_{P_i}(r) \simeq w_{P_i}(0)
	\simeq
	\frac{(1+\nu) P_i}{4\pi R E } 
	\left[
	\frac{-2(1-\nu)-2(1-2\nu)\sin(\theta_i/2)+(7-8\nu)\sin^2(\theta_i/2)}{\sin(\theta_i/2)}
	\right]
	\end{equation}
	Determination of nonlocal radial displacements $u_{P_i}(r)$ is addressed in the next section and it is a major contribution of this work. It is worth noting that if both local and nonlocal radial displacements are assumed negligible, the nonlocal contact formulation proposed by \cite{Gonzalez-2012} is recovered. 
	
	From the above particle deformation analysis, an important conclusion can be drawn. Although physical quantities, such as pressure and contact radius, compatibility conditions and equilibrium are defined in the deformed configuration, it is more convenient to work in the reference configuration. Physical quantities can be easily converted to spacial quantities by using the deformation mapping $\bm{\varphi(\mathbf{X})}$ and push-forward operations. This understanding will be used in the analysis performed in subsequent sections.
	
	\section{Nonlocal effects in the evolution of inter-particle contact area} \label{sec3}
	
	\begin{figure}[t]
		\centering
		\includegraphics[keepaspectratio,scale=0.76]{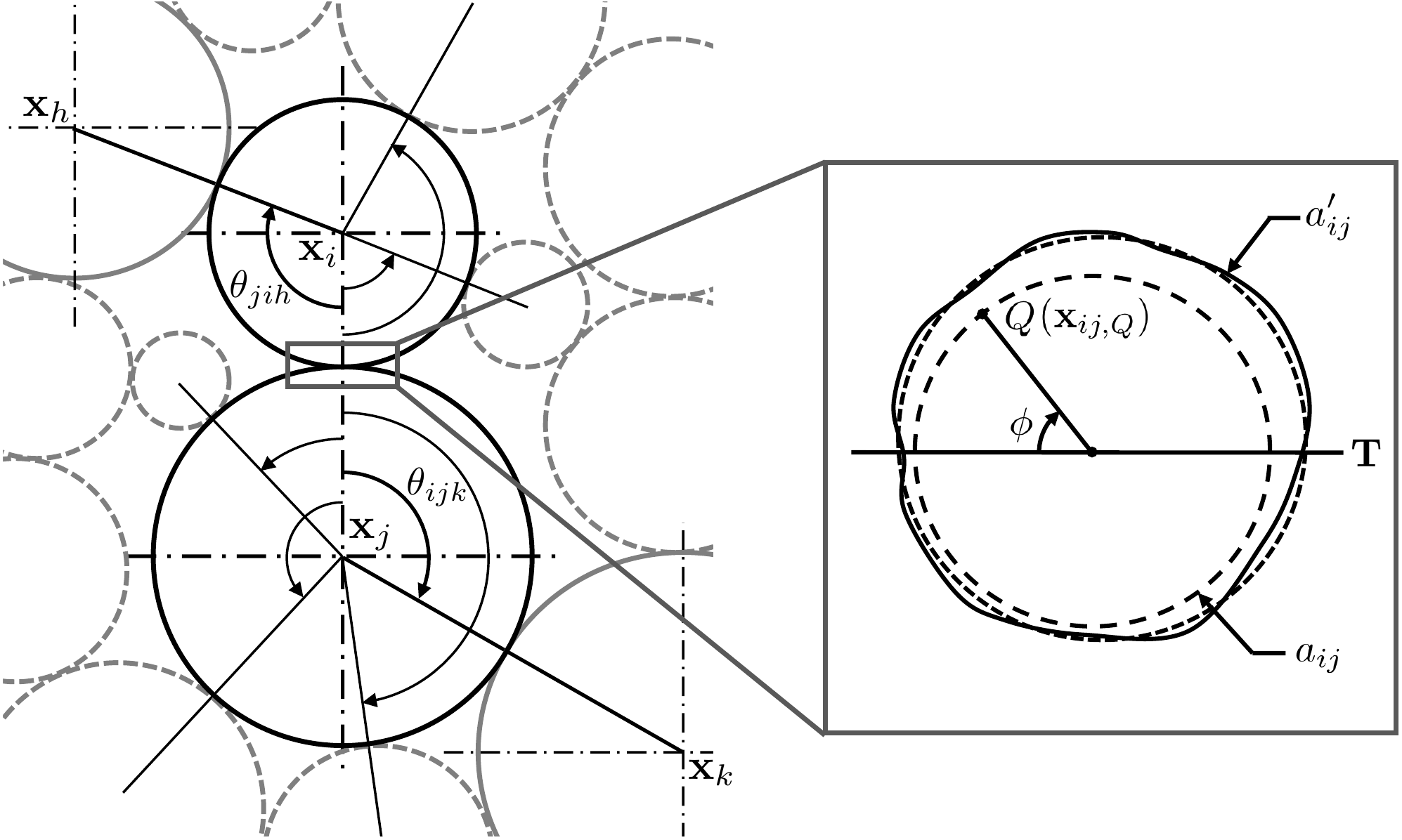}
		\caption{Depiction of nonlocal effects on the contact area between two dissimilar particles being compressed by a general loading configuration. The figure shows a schematic of the loading configuration, and the contact configuration without (long dashed curve) and with (solid curve) nonlocal correction. The contact with effective nonlocal correction is depicted by a short dashed curve, with radius $a'_{ij}$.}
		\label{fig:NL_contact}
	\end{figure}
	
	We now consider the evolution of contact surface between two elastic spherical particles $i$ and $j$ of radius $R_i$ and $R_j$, and material properties $E_i$, $\nu_i$ and $E_j$, $\nu_j$ respectively, being pressed together in a general configuration of particles simulating a confined granular system (Figure \ref{fig:NL_contact}). We propose that the particles deform to accommodate a flat contact surface of effective radius $a'_{ij}$, given by

	\begin{equation} \label{6}
	a'_{ij} = a_{ij} + \frac{1}{2}
		\left[
			\frac{P_{ij}}{m^\mathrm{L}_{ij}(a_{ij})}
			+\sum_{h\in\mathcal{N}_i, h\neq{j}} \frac{P_{ih}}{m^\mathrm{NL}_{jih}(\mathbf{x}_j, \mathbf{x}_i, \mathbf{x}_h)}
			+\sum_{k\in\mathcal{N}_j, k\neq{i}} \frac{P_{jk}}{m^\mathrm{NL}_{ijk}(\mathbf{x}_i, \mathbf{x}_j, \mathbf{x}_k)}\right]
	\end{equation}
	where $a_{ij}$ is the contact radius in the reference configuration, $\mathbf{x}_{i}$ and $\mathcal{N}_{i}$ are the position and neighbors of particle $i$, and $P_{ij}$ is the effective contact force between particles $i$ and $j$. The factor of $1/2$ signifies an average of the radial deformations at the contact due to particles $i$ and $j$ and their respective neighbors.
	
	Term $m^\mathrm{L}_{ij}$ corresponds to the local contribution to radial deformation of the contact boundary. Using eq. \ref{4} for $r = a_{ij}$, it can be expressed as
	\begin{equation} \label{7}
	\frac{1}{m^\mathrm{L}_{ij}} 
	= 
	\sum_{l=i, j}
	\frac{1+\nu_l}{4\pi E_l R_l}
	\left[
	\frac{1}{\sqrt{2}}\sqrt{1+\sqrt{1-\left(\frac{a_{ij}}{R_l}\right)^2}}
	-
	2(1-2\nu_l)\left(\frac{R_l}{a_{ij}}\right)
	\left[1-\frac{1}{\sqrt{2}}\sqrt{1-\sqrt{1-\left(\frac{a_{ij}}{R_l}\right)^2}}\right]
	\right]
	\end{equation}
	
	Term $m^\mathrm{NL}_{jih}$ corresponds to the nonlocal contribution to the radial deformation of the contact boundary, given by
	\begin{equation} \label{8}
	\frac{1}{m^\mathrm{NL}_{jih}} = \frac{1}{2\pi}\int_{0}^{2\pi}\frac{\mathrm{d}\phi}{m^\mathrm{NL}_{jih, Q}(\mathbf{x}_{j}, \mathbf{x}_{i}, \mathbf{x}_{h}, \phi)}
	\end{equation}
	where $m^\mathrm{NL}_{jih,Q}$ corresponds to a particular point $Q$ on the contact boundary, denoted by its angular position $\phi$ with respect to the plane $\mathbf{T}$ given by the equation  $\left[(\mathbf{x}_i-\mathbf{x}_j)\times(\mathbf{x}_i-\mathbf{x}_h)\right]\boldsymbol{\cdot}(\mathbf{x}-\mathbf{x}_i)=\mathbf{0}$ (the plane shown in Figure \ref{fig:NL_contact}). Specifically, $m^\mathrm{NL}_{jih,Q}$ is mathematically represented by	
	\begin{equation} \label{9}
	\begin{aligned}
	\frac{1}{m^\mathrm{NL}_{jih, Q}} &=\frac{1+\nu_i}{2\pi E_iR_i}\left[\frac{\sin\theta_{jih}\cos\phi\left(\sin(\theta_{jih}/2)-\sin(\beta_{jih,Q}/2)\right)\left(\sin(\theta_{jih}/2)\sin(\beta_{jih,Q}/2)-2+2\nu_i\right)}{2\sin(\theta_{jih}/2)\sin(\beta_{jih,Q}/2)}\right. \\
	&\quad\left.+\sqrt{1-\sin^2\theta_{jih}\cos^2\phi}\left[\frac{\cos(\beta_{jih,Q}/2)-\cos(\theta_{jih}/2)}{2}\right .\right. \\
	&\quad\left.\left.-(1-2\nu_i)\left(\frac{1-\sin(\beta_{jih,Q}/2)}{\sin\beta_{jih,Q}}-\frac{\cos(\theta_{jih}/2)}{2\sin(\theta_{jih}/2)(1+\sin(\theta_{jih}/2))}\right)\right]\vphantom{\frac{\sin\theta_{jih}\cos\phi\left(\sin(\theta_{jih}/2)-\sin(\beta_{jih,Q}/2)\right)\left(\frac{1}{2}\sin(\theta_{jih}/2)\sin(\beta_{jih,Q}/2)-1+\nu\right)}{\sin(\theta_{jih}/2)\sin(\beta_{jih,Q}/2)}}\right]
	\end{aligned}
	\end{equation}
	where $\beta_{jih,Q}=\widehat{\mathbf{x}_{ij,Q}\mathbf{x}_i\mathbf{x}	_h}$ is the angle between point $Q$ and position coordinates of particles $i$ and $h$, given by
	\begin{equation} \label{10}
	\beta_{jih, Q} = \cos^{-1}\left[\cos \left|\theta_{jih} - \sin^{-1}\left(\frac{a_{ij}}{R_i}\right)\right| - \left(\frac{a_{ij}}{R_i}\right)(1-\cos \phi)\sin\theta_{jih}\right]
	\end{equation}
	and $\theta_{jih}=\widehat{\mathbf{x}_j\mathbf{x}_i\mathbf{x}_h}$ is the angle between position coordinates of particles $j$, $i$ and $h$. A detailed derivation of $m^\mathrm{NL}_{jih,Q}$ is presented in \ref{appA}. It essentially entails calculation of the radial displacement of point $Q$ on the boundary of the deforming surface due to one of the forces $P_i$ depicted in Figure~\ref{fig:deformation}.
	
	In eq. \ref{8}, $m^\mathrm{NL}_{jih}$ is calculated by taking an average of the radial displacements due to nonlocal forces exerted by particles $h$ on particle $i$ across the contact boundary between particles $i$ and $j$. Such averaging is necessary due to the fact that the nonlocal radial contribution of a particular force is asymmetric over the contact boundary due to dependence on variable angle $\phi$. By averaging over $\phi$, the assumed symmetry in the deformation field is recovered, allowing the deformed contact surface to be approximated by a circular surface, although resulting in a shift in the position of its center which will be neglected in this work (see insert in Figure \ref{fig:NL_contact} and \ref{appA}, eq. \ref{48}). It was found that the integral in eq. \ref{8} does not have an exact closed-form solution; however, its numerical integration is convergent and is used in further analysis.
	
	\begin{figure}[t]
		\centering
		\begin{subfigure}[b]{0.49\linewidth}
			\includegraphics[keepaspectratio,scale=0.67]{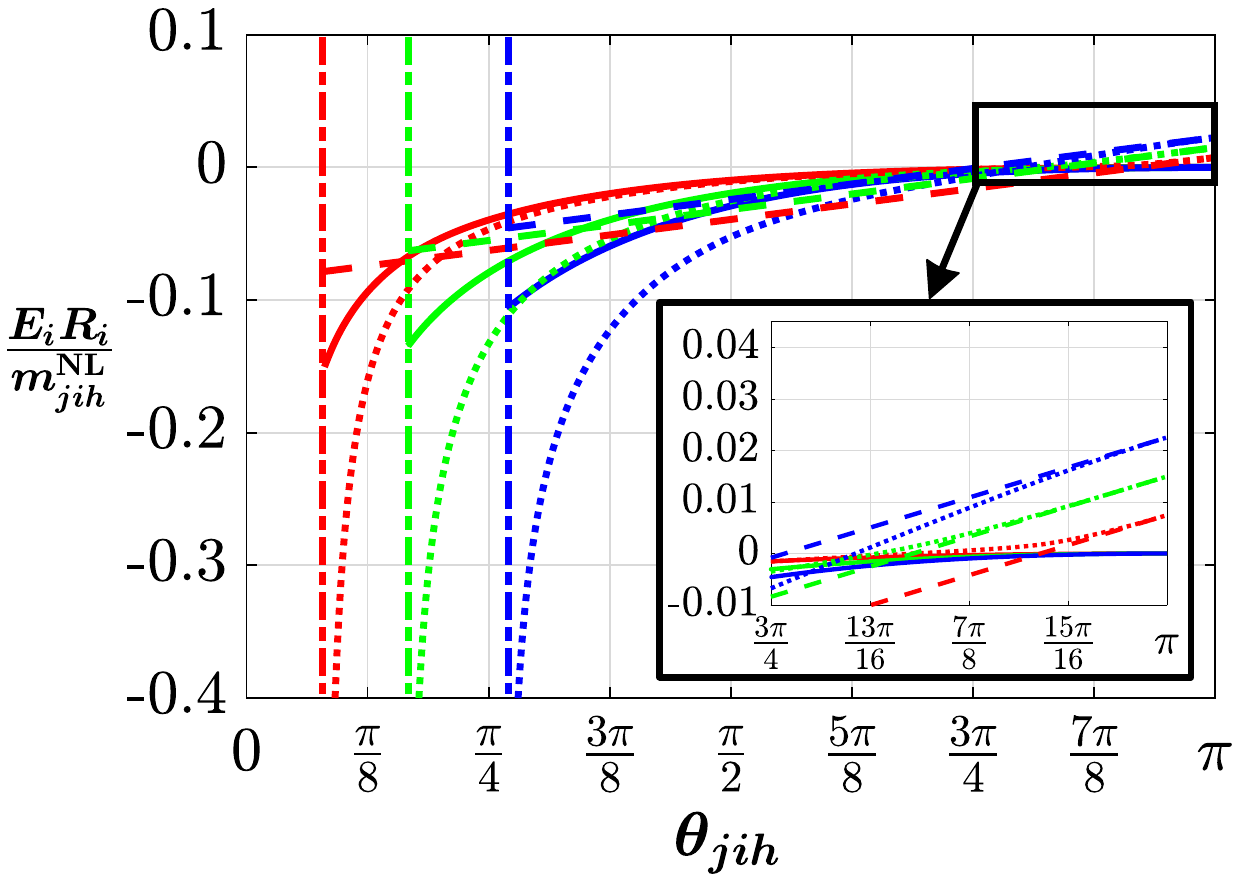}
			\caption{$\nu=0.15$}
			\label{fig:anlyticnumcomp_nu015}
		\end{subfigure}
		\begin{subfigure}[b]{0.49\linewidth}
			\includegraphics[keepaspectratio,scale=0.67]{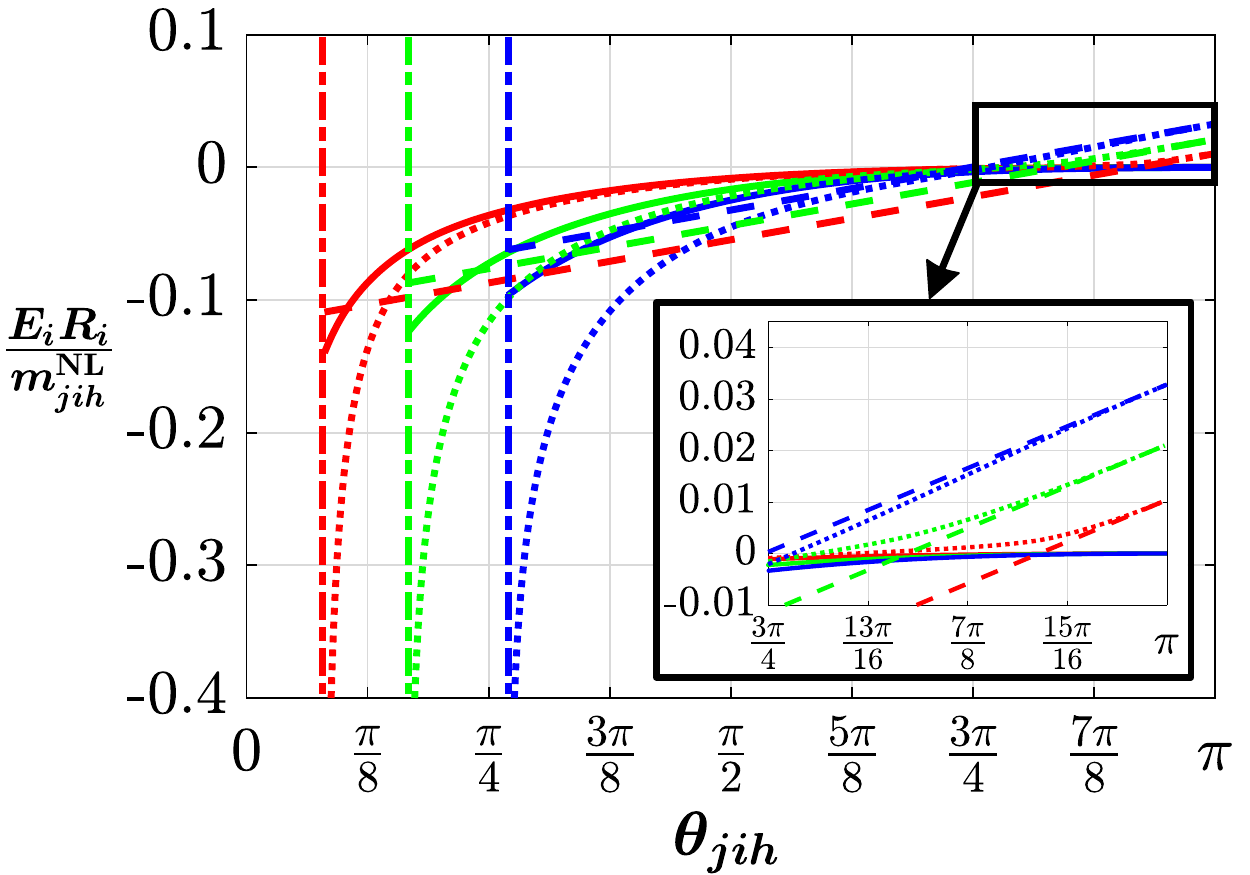}
			\caption{$\nu=0.3$}
			\label{fig:anlyticnumcomp_nu03}
		\end{subfigure}
		\begin{subfigure}[b]{0.49\linewidth}
			\includegraphics[keepaspectratio,scale=0.67]{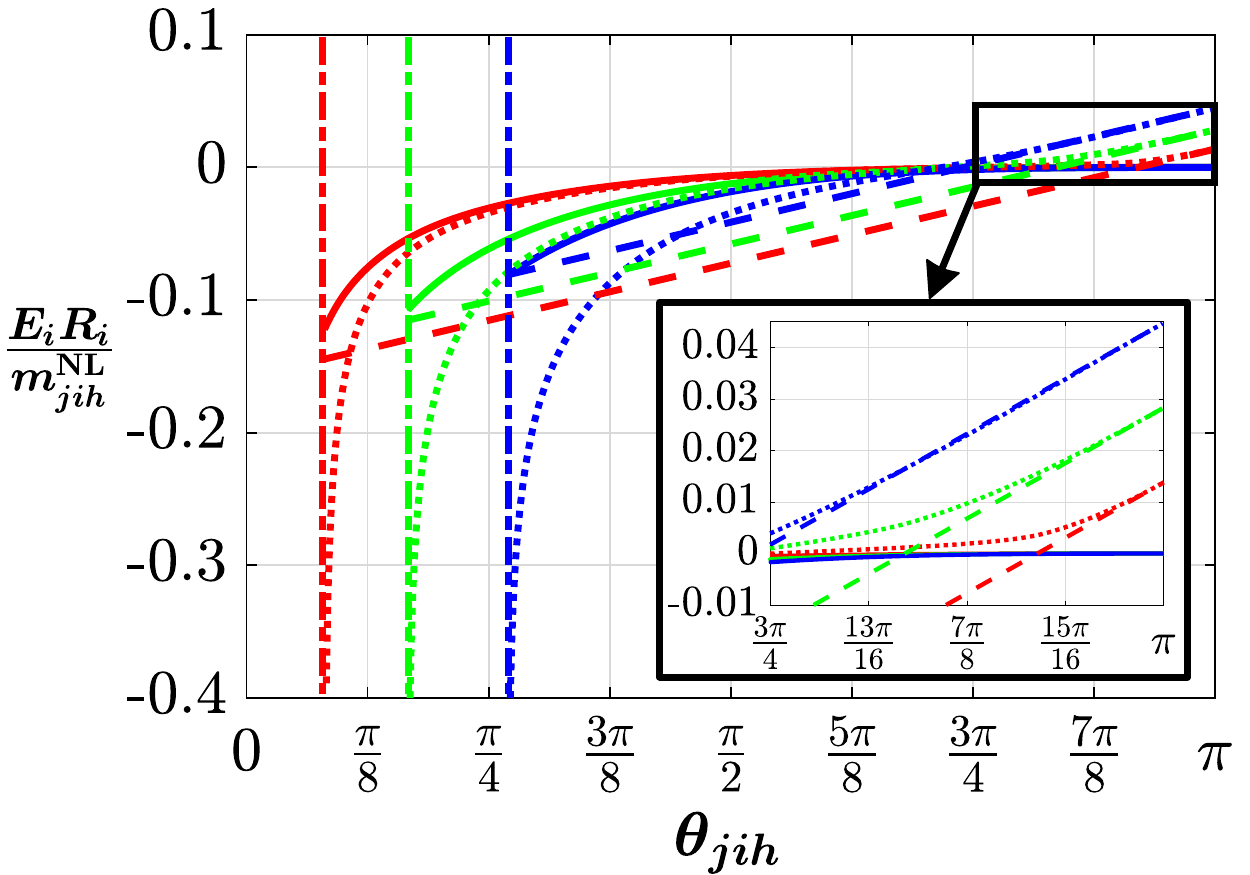}
			\caption{$\nu=0.45$}
			\label{fig:anlyticnumcomp_nu045}
		\end{subfigure}
		\caption{Plots of ${E_i R_i}/{m^{\mathrm{NL}}_{jih}}$ obtained by truncated Taylor expansion at $a_{ij}/R_i = 0$ (eq. \ref{11}, solid curves) and $\theta_{jih}=\pi$ (eq. \ref{12}, dashed curves) versus angular distance $\theta_{jih}$, compared with numerical solution of eq. \ref{8} (dotted curves). Graphs for three different values of $\nu_i$ equal to (\subref{fig:anlyticnumcomp_nu015}). 0.15, (\subref{fig:anlyticnumcomp_nu03}).  0.3 and (\subref{fig:anlyticnumcomp_nu045}). 0.45 are presented, each having plots evaluated at three different values of $a_{ij}/R_i$ ($\text{red curves for }a_{ij}/R_i=0.25\text{, green curves for }a_{ij}/R_i=0.5\text{ and blue curves for }a_{ij}/R_i=0.75$). The lower bound of $\theta_{jih} = \sin^{-1}(a_{ij}/R_i)$ (dashed-dotted lines) for evaluation of nonlocal contributions has also been delineated in each of the graphs.}
		\label{fig:anlyticnumcomp1}
	\end{figure}

	We next proceed to identify a closed-form approximate solution for $m^\mathrm{NL}_{jih}$ amenable to a computationally tractable implementation. To this end, we first carry out a Taylor series expansion of $1/{m^\mathrm{NL}_{jih,Q}}$ about $a_{ij}/R_i=0$ to obtain 
	\begin{equation} \label{11}
		\frac{1}{2\pi}\int_{0}^{2\pi}\frac{\mathrm{d}\phi}{m^\mathrm{NL}_{jih, Q}(\mathbf{x}_{j}, \mathbf{x}_{i}, \mathbf{x}_{h}, \phi)}
		=
		\frac{1+\nu_i}{16\pi E_iR_i}
		\left[\frac{\cos^2\left(\theta_{jih}/2\right)(4\nu_i-3-\cos\theta_{jih})}{\sin\left(\theta_{jih}/2\right)}\right]
		\left(\frac{a_{ij}}{R_i}\right)
		+ 
		\mathcal{O}\left(\left(\frac{a_{ij}}{R_i}\right)^2\right)
	\end{equation}	
	Similary, we carry out a Taylor expansion of $1/{m^\mathrm{NL}_{jih,Q}}$ about  $\theta_{jih}=\pi$ to obtain 
	\begin{equation} \label{12}
		\frac{1}{2\pi}\int_{0}^{2\pi}\frac{\mathrm{d}\phi}{m^\mathrm{NL}_{jih, Q}(\mathbf{x}_{j}, \mathbf{x}_{i}, \mathbf{x}_{h}, \phi)}
		= 
		\frac{1}{m^\mathrm{NL}_{jih,\pi}}
		-
		\frac{(1+\nu_i)(1+2\nu_i)}{16\pi E_iR_i}(\pi-\theta_{jih})
		+
		\mathcal{O}\left((\pi-\theta_{jih})^2\right)
	\end{equation}
	where $m^\mathrm{NL}_{jih,\pi}$ corresponds to the analytical solution of eq. \ref{8} for $\theta_{jih}=\pi$, that is
	\begin{equation} \label{13}
		\frac{1}{m^\mathrm{NL}_{jih,\pi}} 
		=
		\frac{1+\nu_i}{4\pi E_i R_i}
		\left[\frac{1}{\sqrt{2}}\sqrt{1-\sqrt{1-\left(\frac{a_{ij}}{R_i}\right)^2}}
		-
		2(1-2\nu_i)
		\left(\frac{R_i}{a_{ij}}\right)
		\left[1-\frac{1}{\sqrt{2}}\sqrt{1+\sqrt{1-\left(\frac{a_{ij}}{R_i}\right)^2}}\right]\right]
	\end{equation}	
	Figure \ref{fig:anlyticnumcomp1} presents a comparison of the two approximate analytical solutions obtained from truncated Taylor expansions at $a_{ij}/R_i = 0$ (eq. \ref{11}) and $\theta_{jih}=\pi$ (eq. \ref{12}) with the numerical solution of eq. \ref{8}, by plotting the dimensionless quantity ${E_i R_i}/{m^{\mathrm{NL}}_{jih}}$ with respect to $\theta_{jih}$ for all possible values of $\theta_{jih}\in\left(\sin^{-1}\left(a_{ij}/R_i\right),\pi\right]$. To gain a better insight into the non-linear dependency of the expressions on $a_{ij}/R_i$ and $\nu_i$, graphs for three different values of $\nu_i$ (= 0.15, 0.3 and 0.45) are shown, with each graph having plots evaluated at three different values of $a_{ij}/R_i$ (= 0.25, 0.5 and 0.75). Analysis of the plots suggests that the numerical solution is fairly represented by eq. \ref{11} until it shifts to positive values near $\theta_{jih}={3\pi/4}$. Thereafter, the numerical solution is well represented by eq. \ref{12}. Accordingly, we propose a piecewise continuous function to represent $m^\mathrm{NL}_{jih}$, given by

	\begin{figure}[t]
		\centering
		\begin{subfigure}[b]{0.49\linewidth}
			\includegraphics[keepaspectratio,scale=0.67]{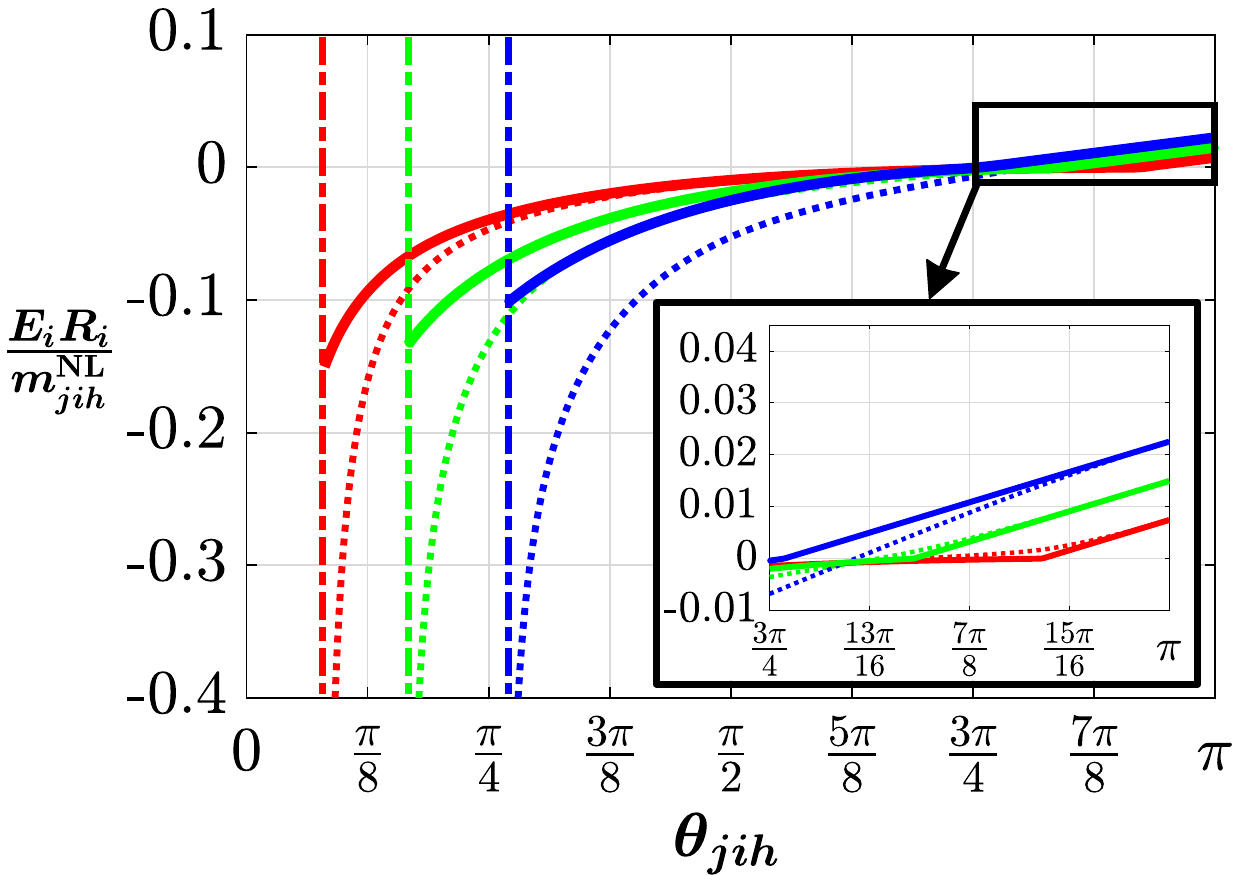}
			\caption{$\nu=0.15$}
			\label{fig:anlyticnumcomp2_nu015}
		\end{subfigure}
		\begin{subfigure}[b]{0.49\linewidth}
			\includegraphics[keepaspectratio,scale=0.67]{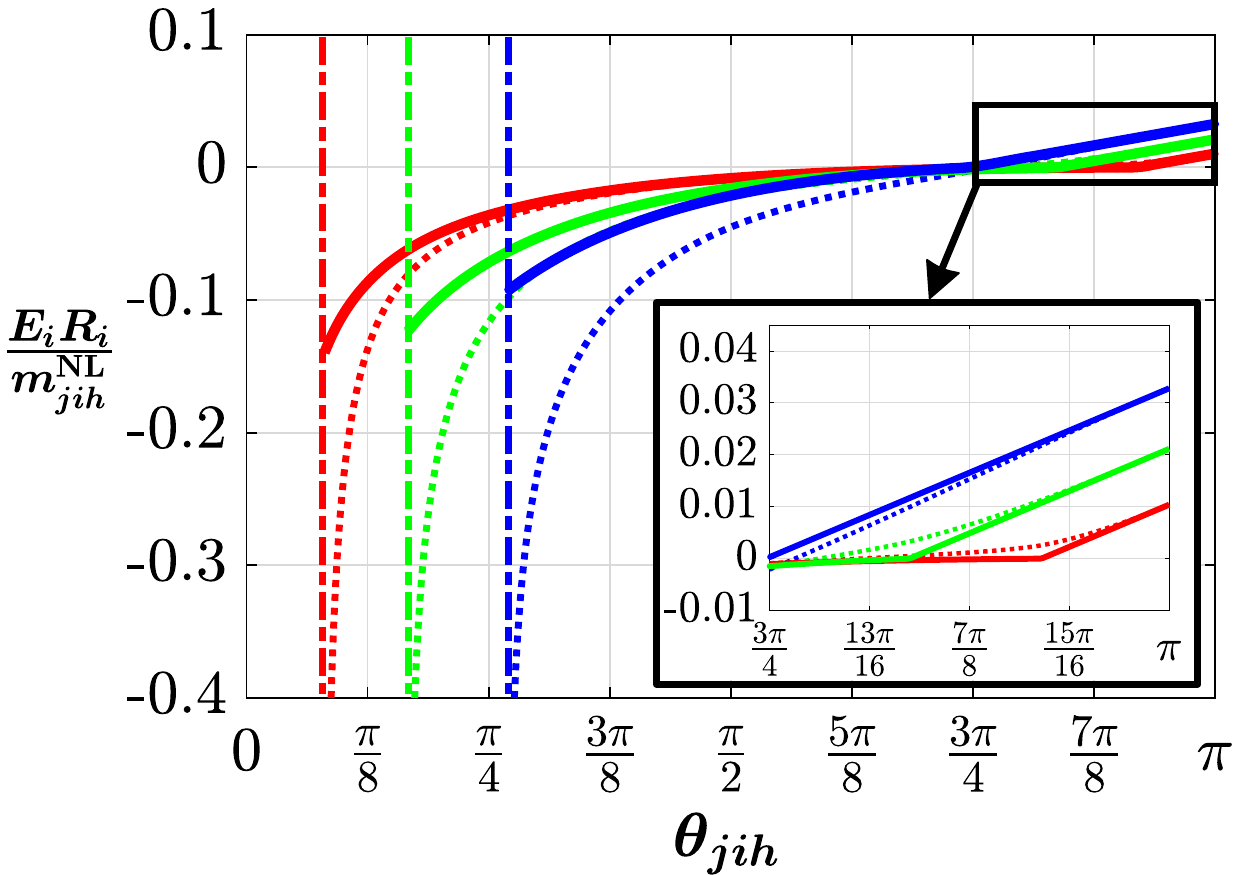}
			\caption{$\nu=0.3$}
			\label{fig:anlyticnumcomp2_nu03}
		\end{subfigure}
		\begin{subfigure}[b]{0.49\linewidth}
			\includegraphics[keepaspectratio,scale=0.67]{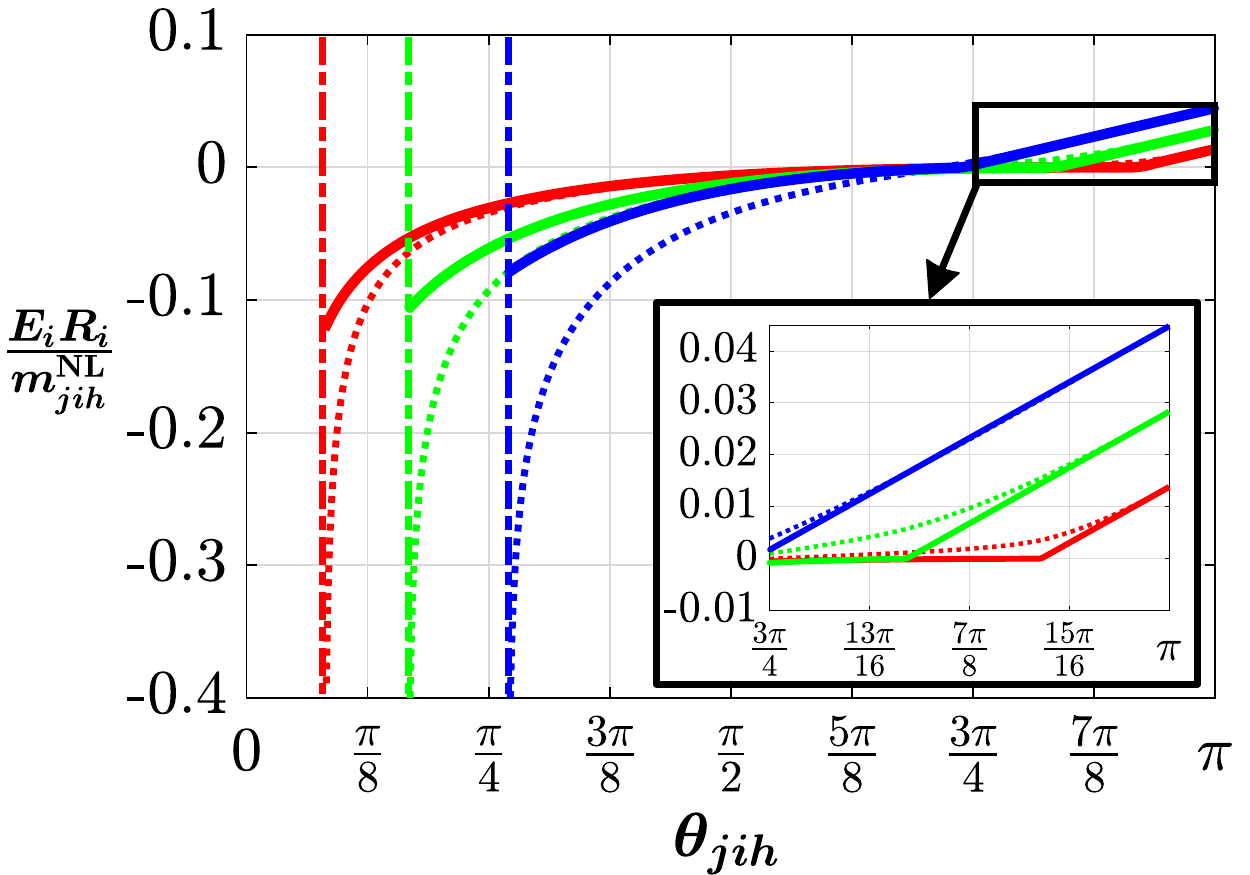}
			\caption{$\nu=0.45$}
			\label{fig:anlyticnumcomp2_nu045}
		\end{subfigure}
		\caption{Plots of ${E_i R_i}/{m^{\mathrm{NL}}_{jih}}$ obtained by piecewise approximate solution eq. \ref{14} (bold solid curves) versus angular distance $\theta_{jih}$, compared with numerical solution of eq. \ref{8} (dotted curves). The values of $\nu_i$ and $a_{ij}/R_i$ are the same as in figure \ref{fig:anlyticnumcomp1}.}
		\label{fig:anlyticnumcomp2}
	\end{figure}
	\begin{equation} \label{14}
	\dfrac{1}{m^\mathrm{NL}_{jih}} 
	\simeq \small
	\left\{ \begin{array}{ll}
	\dfrac{1+\nu_i}{16\pi E_i R_i}\left[\dfrac{\cos^2(\theta_{jih}/2)(4\nu_i-3-\cos\theta_{jih})}{\sin(\theta_{jih}/2)}-\dfrac{\cos^2(\theta^\mathrm{c}_{jih}/2)(4\nu_i-3-\cos\theta^\mathrm{c}_{jih})}{\sin(\theta^\mathrm{c}_{jih}/2)}\right]\left(\dfrac{a_{ij}}{R_i}\right)&\mbox{$\theta_{jih} \leq \theta^\mathrm{c}_{jih}$} \\ \\
	\dfrac{1}{m^\mathrm{NL}_{jih,\pi}}-\dfrac{(1+\nu_i)(1+2\nu_i)}{16\pi E_i R_i}(\pi-\theta_{jih})&\mbox{$\theta_{jih} > \theta^\mathrm{c}_{jih}$}\end{array} \right.
	\end{equation} 
	\normalsize where $\theta^\mathrm{c}_{jih}$ is the critical value of angular distance at which $1/m^{\mathrm{NL}}_{jih}=0$, given by
	\begin{equation} \label{15}
	\theta^\mathrm{c}_{jih} = \pi-\frac{16\pi E_i R_i}{(1+\nu_i)(1+2\nu_i)m^\mathrm{NL}_{jih,\pi}}
	\end{equation}
	Figure \ref{fig:anlyticnumcomp2} shows a comparison of the numerical solution of $1/m^\mathrm{NL}_{jih}$ and the proposed piecewise continuous approximate solution, using the same values of $a_{ij}/R_i$ and $\nu_i$ as in Figure \ref{fig:anlyticnumcomp1}. The accuracy of the closed-form approximate solution is acceptable and the computational tractability of the formulation is attained.
	
	\section{Curvature correction to the  nonlocal contact formulation} \label{sec4}
	We again consider a configuration of two elastic spheres $i$ and $j$ being pressed together along the direction $\mathbf{z}_{ij}=(\mathbf{x}_i-\mathbf{x}_j)/||\mathbf{x}_i-\mathbf{x}_j||$ by a general configuration of concentrated forces, as depicted in Figure \ref{fig:NL_contact}. According to the nonlocal contact formulation \citep*{Gonzalez-2012}, any point within the contact area satisfies the following compatibility equation
	\begin{equation} \label{16}
	\gamma_{ij} = R_i - \sqrt{R^2_i - r^2_{ij}} + R_j - \sqrt{R^2_j - r^2_{ij}} + w_i(r_{ij}) + w_j(r_{ij}) - \gamma^{\mathrm{NL}}_{ij} 
	\end{equation}
	where $\gamma_{ij}$ is the relative displacement of $\mathbf{x}_i$ and $\mathbf{x}_j$ along the direction $\mathbf{z}_{ij}$, $r_{ij}$ is the radial coordinate from axis $\mathbf{z}_{ij}$ of surface points in the reference configuration, and $w_i(r_{ij})$ and $w_j(r_{ij})$ are the vertical displacements of surface points located at $r_{ij}$ on spheres $i$ and $j$. The last term in the equation, $\gamma^{\mathrm{NL}}_{ij}$, is the total nonlocal contribution to vertical displacements induced by all neighbors of the two particles, i.e.,
	\begin{equation} \label{17}
	\gamma^{\mathrm{NL}}_{ij} 
	=
	\sum_{h\in\mathcal{N}_i, h\neq{j}} \frac{P_{ih}}{n^\mathrm{NL}_{jih}(\mathbf{x}_j, \mathbf{x}_i, \mathbf{x}_h)}
	+
	\sum_{k\in\mathcal{N}_j, k\neq{i}} \frac{P_{jk}}{n^\mathrm{NL}_{ijk}(\mathbf{x}_i, \mathbf{x}_j, \mathbf{x}_k)}
	\end{equation}
	with $n^\mathrm{NL}_{jih}$ derived from eq. \ref{5}  \citep{Gonzalez-2012}
	\begin{equation} \label{18}
	\frac{1}{n^\mathrm{NL}_{jih}}
	=
	\frac{(1+\nu_i)}{4\pi R_i E_i } 
	\left[
	\frac{-2(1-\nu_i)-2(1-2\nu_i)\sin(\theta_{jih}/2)+(7-8\nu_i)\sin^2(\theta_{jih}/2)}{\sin(\theta_{jih}/2)}
	\right]
	\end{equation}
	An important assumption of the Hertz theory is that the profile of the undeformed spherical contact surface is replaced by the first term of its Taylor series expansion, i.e.
	\begin{equation} \label{19}
	R_i - \sqrt{R^2_i - r^2_{ij}} = R_i\left[\frac{r^2_{ij}}{2R^2_i}+\mathcal{O}\left(\frac{a^4_{ij}}{R^4_i}\right)\right]\simeq\frac{r^2_{ij}}{2R_i}
	\end{equation}
	This approximation largely deviates from the exact solution for moderate to high mesoscopic deformations. However, the error can be controlled by including more terms in the series to further correct the profile curvature; for example, the first four terms are given by
	\begin{equation} \label{20}
	R_i - \sqrt{R^2_i - r^2_{ij}} = R_i\left[\frac{r^2_{ij}}{2R^2_i}+ \frac{r^4_{ij}}{8R^4_i} + \frac{r^6_{ij}}{16R^6_i}+\mathcal{O}\left(\frac{a_{ij}^8}{R^8_i}\right)\right]\simeq\frac{r_{ij}^2}{2R_i} + \frac{r^4_{ij}}{8R^3_i} + \frac{r^6_{ij}}{16R^5_i} + \frac{5r^8_{ij}}{128R^7_i} 
	\end{equation}
	Therefore, by adopting a four-term curvature correction, eq. \ref{16} simplifies to
	\begin{equation} \label{21}
	w_i(r_{ij}) + w_j(r_{ij}) = (\gamma_{ij}+ \gamma^{\mathrm{NL}}_{ij}) - \frac{r^2_{ij}\mathbb{A}_{ij}}{2} - \frac{r^4_{ij}\mathbb{B}_{ij}}{8} -\frac{r^6_{ij}\mathbb{C}_{ij}}{16} - \frac{5r^8_{ij}\mathbb{D}_{ij}}{128}  
	\end{equation}   
	where
	\begin{equation*}
	\mathbb{A}_{ij} = \frac{1}{R_i} + \frac{1}{R_j} \hspace{20pt} \mathbb{B}_{ij}=\frac{1}{R^3_i}+\frac{1}{R^3_j} \hspace{20pt} \mathbb{C}_{ij}=\frac{1}{R^5_i}+\frac{1}{R^5_j} \hspace{20pt} \mathbb{D}_{ij}=\frac{1}{R^7_i}+\frac{1}{R^7_j}
	\end{equation*}
	
\begin{figure}[t]
		\centering
		\includegraphics[keepaspectratio,scale=0.7]{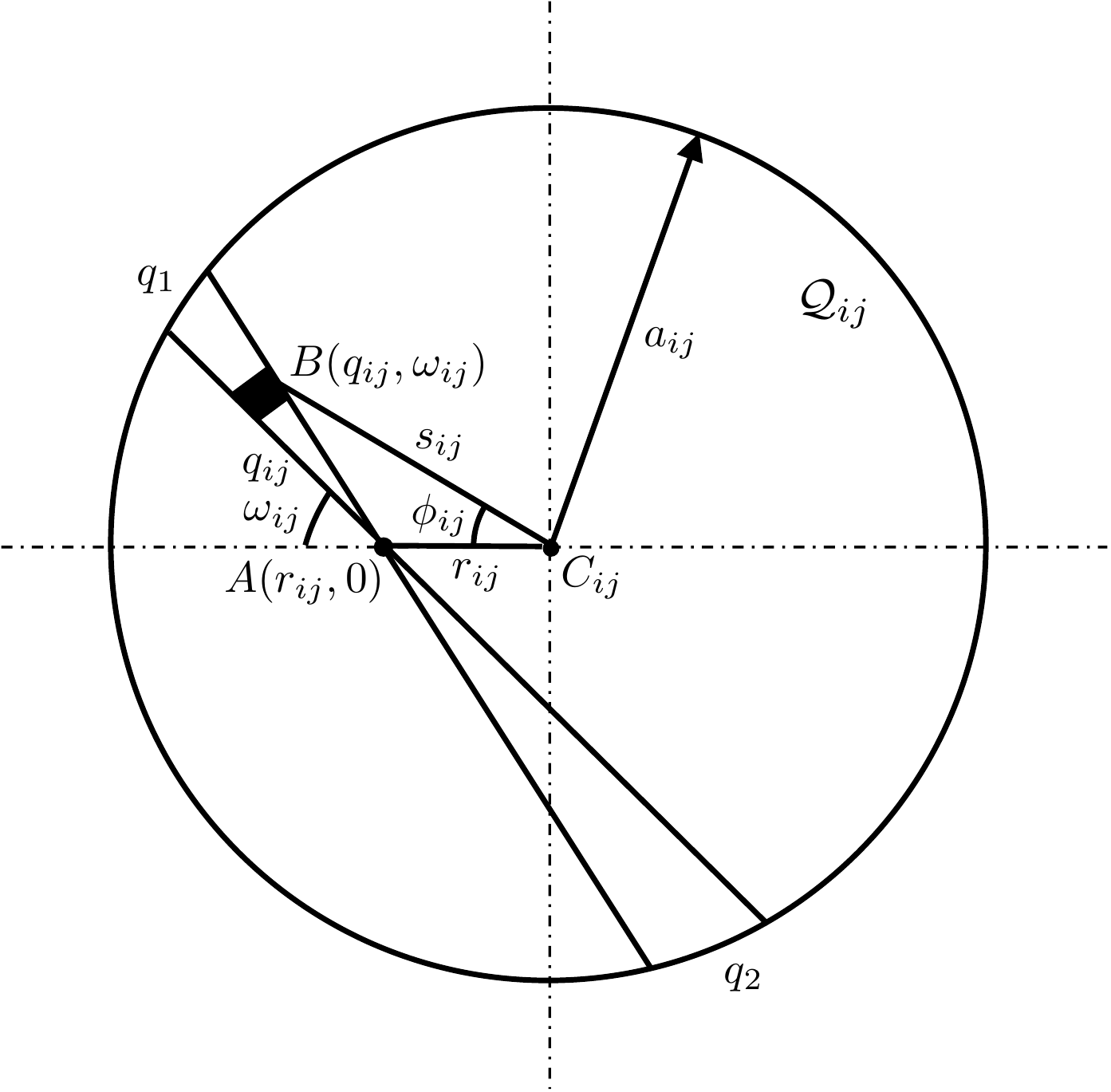}
		\caption{Depiction of the circular contact region $\mathcal{Q}_{ij}$ of contact radius $a_{ij}$ under the action of a distributed contact pressure, with $B(q_{ij}, \omega_{ij})$ being an elemental region on which the pressure distribution is considered.}
		\label{fig:pressureprofile}
	\end{figure}

	Next, we determine the pressure distribution $p_{ij}$ over the circular contact region $\mathcal{Q}_{ij}$ of contact radius $a_{ij}$ in the reference configuration, depicted in Figure \ref{fig:pressureprofile}, which is compatible with eq. \ref{21}. Specifically, for a pressure $p_{ij}(q_{ij},\omega_{ij})$ acting over an elemental region $B$ of area $q_{ij}\,dq_{ij}\,d\omega_{ij}$, the vertical displacement field according to the theory of elasticity \cite[pg. 53]{Johnson-1985} is given by
	\begin{equation} \label{22}
	w_i(r_{ij})+w_j(r_{ij})
	=
	\left(\frac{1-\nu^2_i}{\pi{E_i}}+\frac{1-\nu^2_j}{\pi{E_j}}\right)\int\int_{\mathcal{Q}_{ij}}^{}p_{ij}(q_{ij},\omega_{ij})dq_{ij}d\omega_{ij}
	\end{equation}
	and thus, using eq. \ref{22} in \ref{21}, we get
	\begin{equation} \label{23}
	\left(\frac{1-\nu^2_i}{\pi{E_i}}+\frac{1-\nu^2_j}{\pi{E_j}}\right)\int\int_{\mathcal{Q}_{ij}}^{}p_{ij}(q_{ij},\omega_{ij})dq_{ij}d\omega_{ij} = (\gamma_{ij}+ \gamma^{\mathrm{NL}}_{ij}) - \frac{r^2_{ij}\mathbb{A}_{ij}}{2} - \frac{r^4_{ij}\mathbb{B}_{ij}}{8} -\frac{r^6_{ij}\mathbb{C}_{ij}}{16} - \frac{5r^8_{ij}\mathbb{D}_{ij}}{128}  
	\end{equation}
	
	The pressure distribution $p_{ij}$ is readily available by following the method reported by \cite{LUO-1958}, which involves taking an approximate form of $p_{ij}$ to solve the integral, followed by comparing coefficients of like powers of $r_{ij}$ on both sides of the equation. For reference, the solution method is described in \ref{appB}. Using the pressure distribution functions, contact radius-force-displacement relationships for two-term, three-term and four-term curvature corrections are derived in the following sub-sections.

	\subsection{Two-term curvature correction} \label{sec4.1}
	
	\noindent\textit{Pressure distribution} (\ref{secB.1}, eq. \ref{72}):
	\begin{equation} \label{24}
	p_{ij}(r_{ij}) = \frac{2a_{ij}}{\pi}\left(\frac{1-\nu^2_i}{{E_i}}+\frac{1-\nu^2_j}{{E_j}}\right)^{-1}\left(1-\frac{r^2_{ij}}{a^2_{ij}}\right)^{1/2} \left[\mathbb{A}_{ij}+\frac{2a^2_{ij}\mathbb{B}_{ij}}{9}
	\left(1+2\frac{r^2_{ij}}{a^2_{ij}}\right)\right]
	\end{equation}
	\textit{Contact radius-displacement relationship}:
	according to (\ref{secB.1}, eq. \ref{71})
	\begin{equation} \label{25}
	a_{ij}^4\left(\frac{\mathbb{B}_{ij}}{3}\right) + a^2_{ij}\mathbb{A}_{ij} - (\gamma_{ij}+\gamma^{\mathrm{NL}}_{ij}) = 0
	\end{equation}
	which yields only one real and positive solution given by
	\begin{equation} \label{26}
	a_{ij} = \left[\frac{3}{2\mathbb{B}_{ij}}\left(\sqrt{\frac{4\mathbb{B}_{ij}}{3}(\gamma_{ij}+\gamma^{\mathrm{NL}}_{ij})+\mathbb{A}^2_{ij}}-\mathbb{A}_{ij}\right)\right]^{1/2}
	\end{equation}
	\textit{Contact force-radius-displacement relationship}:
	\begin{equation} \label{27}
	P_{ij} =
	\int_{0}^{a_{ij}}p_{ij}(r_{ij})2\pi{r_{ij}}dr_{ij} =
	\frac{4E^*_{ij}}{3}a_{ij}(\gamma_{ij}+\gamma^{\mathrm{NL}}_{ij})^3\left[\mathbb{A}_{ij}+\frac{2a_{ij}(\gamma_{ij}+\gamma^{\mathrm{NL}}_{ij})^2\mathbb{B}_{ij}}{5}\right]
	\end{equation}
	where
	\begin{equation*}
	\frac{1}{E^*_{ij}} = \frac{1-\nu^2_i}{{E_i}}+\frac{1-\nu^2_j}{{E_j}}
	\end{equation*}
	and $a_{ij}(\gamma_{ij}+\gamma^{\mathrm{NL}}_{ij})$ is given by eq. \ref{26}.

	\subsection{Three-term curvature correction} \label{sec4.2}
	 
	\noindent\textit{Pressure distribution}(\ref{secB.2}, eq. \ref{82}):
	\begin{equation} \label{28}
	p_{ij}(r_{ij}) = \frac{2a_{ij}}{\pi}\left(\frac{1-\nu^2_i}{{E_i}}+\frac{1-\nu^2_j}{{E_j}}\right)^{-1}\left(1-\frac{r^2_{ij}}{a^2_{ij}}\right)^{1/2} \left[\mathbb{A}_{ij}+\frac{2a^2_{ij}\mathbb{B}_{ij}}{9}
	\left(1+2\frac{r^2_{ij}}{a^2_{ij}}\right)+\frac{a^4_{ij}\mathbb{C}_{ij}}{25}\left(3+4\frac{r^2_{ij}}{a^2_{ij}}+8\frac{r^4_{ij}}{a^4_{ij}}\right)\right]
	\end{equation}
	\textit{Contact radius-displacement relationship}:
	according to (\ref{secB.2}, eq. \ref{81})
	\begin{equation} \label{29}
	\gamma_{ij}+\gamma^{\mathrm{NL}}_{ij} = a^2_{ij}\mathbb{A}_{ij}+\frac{a^4_{ij}}{3}\mathbb{B}_{ij}+\frac{a^6_{ij}}{5}\mathbb{C}_{ij}
	\end{equation}
	which yields only one real and positive solution given by
	\begin{equation} \label{30}
	a_{ij}=\left[\frac{\left[(\mathbb{Q}^2_{ij}+4\mathbb{R}^3_{ij})^{1/2}+\mathbb{Q}_{ij}\right]^{1/3}}{9(2)^{1/3}\mathbb{C}_{ij}}-\frac{(2)^{1/3}\mathbb{R}_{ij}}{9\mathbb{C}_{ij}\left[(\mathbb{Q}^2_{ij}+4\mathbb{R}^3_{ij})^{1/2}+\mathbb{Q}_{ij}\right]^{1/3}}-\frac{5\mathbb{B}_{ij}}{9\mathbb{C}_{ij}}\right]^{1/2}
	\end{equation}
	where
	\begin{equation*}
	\begin{aligned}
	\mathbb{Q}_{ij}&=2025\mathbb{A}_{ij}\mathbb{B}_{ij}\mathbb{C}_{ij}-250\mathbb{B}^3_{ij}+3645\mathbb{C}^2_{ij}(\gamma_{ij}+\gamma^{\mathrm{NL}}_{ij}) \\
	\mathbb{R}_{ij}&=135\mathbb{A}_{ij}\mathbb{C}_{ij}-25\mathbb{B}^2_{ij}
	\end{aligned}
	\end{equation*}
	\textit{Contact force-radius-displacement relationship}:
	\begin{equation} \label{31}
	P_{ij} = \frac{4E^*_{ij}}{3}a_{ij}(\gamma_{ij}+\gamma^{\mathrm{NL}}_{ij})^3\left[\mathbb{A}_{ij}+\frac{2a_{ij}(\gamma_{ij}+\gamma^{\mathrm{NL}}_{ij})^2\mathbb{B}_{ij}}{5}+\frac{9a_{ij}(\gamma_{ij}+\gamma^{\mathrm{NL}}_{ij})^4\mathbb{C}_{ij}}{35}\right]
	\end{equation}
	where $a_{ij}(\gamma_{ij}+\gamma^{\mathrm{NL}}_{ij})$ is given by eq. \ref{30}.

	\subsection{Four-term curvature correction} \label{sec4.3} 
	
	\noindent\textit{Pressure distribution} (\ref{secB.3}, eq. \ref{94}):
	\begin{equation} \label{32}
	\begin{aligned}
	p_{ij}(r_{ij}) &= \frac{2a_{ij}}{\pi}\left(\frac{1-\nu^2_i}{{E_i}}+\frac{1-\nu^2_j}{{E_j}}\right)^{-1}\left(1-\frac{r^2_{ij}}{a^2_{ij}}\right)^{1/2} \left[\mathbb{A}_{ij}+\frac{2a^2_{ij}\mathbb{B}_{ij}}{9}
	\left(1+2\frac{r^2_{ij}}{a^2_{ij}}\right)+\frac{a^4_{ij}\mathbb{C}_{ij}}{25}\left(3+4\frac{r^2_{ij}}{a^2_{ij}}+8\frac{r^4_{ij}}{a^4_{ij}}\right)\right.\\
	&\quad\left.+\frac{4a^6_{ij}\mathbb{D}_{ij}}{245}\left(5+6\frac{r^2_{ij}}{a^2_{ij}}+8\frac{r^4_{ij}}{a^4_{ij}}+16\frac{r^6_{ij}}{a^6_{ij}}\right)\right]
	\end{aligned}
	\end{equation}
	\textit{Contact radius-displacement relationship}:
	according to (\ref{secB.3}, eq. \ref{93})
	\begin{equation} \label{33}
	\gamma_{ij}+\gamma^{\mathrm{NL}}_{ij} = a^2_{ij}\mathbb{A}_{ij}+\frac{a^4_{ij}}{3}\mathbb{B}_{ij}+\frac{a^6_{ij}}{5}\mathbb{C}_{ij}+\frac{a^8_{ij}}{7}\mathbb{D}_{ij}
	\end{equation}
	which yields only one real and positive solution given by
	\begin{equation} \label{34}
	a_{ij}=\left[\frac{1}{2}\sqrt{\frac{147\mathbb{C}^2_{ij}}{100\mathbb{D}^2_{ij}}-\frac{14\mathbb{B}_{ij}}{3\mathbb{D}_{ij}}-\mathbb{S}_{ij}-\frac{1}{4\sqrt{\mathbb{S}_{ij}}}\left(\frac{196\mathbb{B}_{ij}\mathbb{C}_{ij}}{15\mathbb{D}^2_{ij}}-\frac{343\mathbb{C}^3_{ij}}{125\mathbb{D}^3_{ij}}-\frac{56\mathbb{A}_{ij}}{\mathbb{D}_{ij}}\right)}-\frac{\sqrt{\mathbb{S}_{ij}}}{2}-\frac{7\mathbb{C}_{ij}}{20\mathbb{D}_{ij}}\right]^{1/2}
	\end{equation}
	where
	\begin{equation*}
	\begin{aligned}
	\mathbb{S}_{ij}&=\frac{7(2)^{1/3}\mathbb{S}^{(1)}_{ij}}{9\mathbb{D}_{ij}\left[\mathbb{S}^{(2)}_{ij}+\sqrt{{\mathbb{S}^{(2)}_{ij}}^2-171500{\mathbb{S}^{(1)}_{ij}}^3}\right]^{1/3}}+\frac{\left[\mathbb{S}^{(2)}_{ij}+\sqrt{{\mathbb{S}^{(2)}_{ij}}^2-171500{\mathbb{S}^{(1)}_{ij}}^3}\right]^{1/3}}{45(2)^{1/3}\mathbb{D}_{ij}}+\frac{49\mathbb{C}^2_{ij}}{100\mathbb{D}^2_{ij}}-\frac{14\mathbb{B}_{ij}}{9\mathbb{D}_{ij}} \\
	\mathbb{S}^{(1)}_{ij}&=35\mathbb{B}^2_{ij}-189\mathbb{A}_{ij}\mathbb{C}_{ij}-540\mathbb{D}_{ij}(\gamma_{ij}+\gamma^{\mathrm{NL}}_{ij}) \\
	\mathbb{S}^{(2)}_{ij}&=4465125\mathbb{A}^2_{ij}\mathbb{D}_{ij}-694575\mathbb{A}_{ij}\mathbb{B}_{ij}\mathbb{C}_{ij}+85750\mathbb{B}^3_{ij}+(3969000\mathbb{B}_{ij}\mathbb{D}_{ij}-1250235\mathbb{C}^2_{ij})(\gamma_{ij}+\gamma^{\mathrm{NL}}_{ij})
	\end{aligned}
	\end{equation*}
	\textit{Contact force-radius-displacement relationship}:
	\begin{equation} \label{35}
	P_{ij} = \frac{4E^*_{ij}}{3}a_{ij}(\gamma_{ij}+\gamma^{\mathrm{NL}}_{ij})^3\left[\mathbb{A}_{ij}+\frac{2a_{ij}(\gamma_{ij}+\gamma^{\mathrm{NL}}_{ij})^2\mathbb{B}_{ij}}{5}+\frac{9a_{ij}(\gamma_{ij}+\gamma^{\mathrm{NL}}_{ij})^4\mathbb{C}_{ij}}{35}+\frac{4a_{ij}(\gamma_{ij}+\gamma^{\mathrm{NL}}_{ij})^6\mathbb{D}_{ij}}{21}\right]
	\end{equation} 
	where $a_{ij}(\gamma_{ij}+\gamma^{\mathrm{NL}}_{ij})$ is given by eq. \ref{34}.
	
	\section{Validation of contact radius and curvature corrections} \label{sec5}
	
\begin{figure}[p]
		\centering
		\begin{subfigure}[b]{0.48\linewidth}
			\includegraphics[keepaspectratio,scale=0.27]{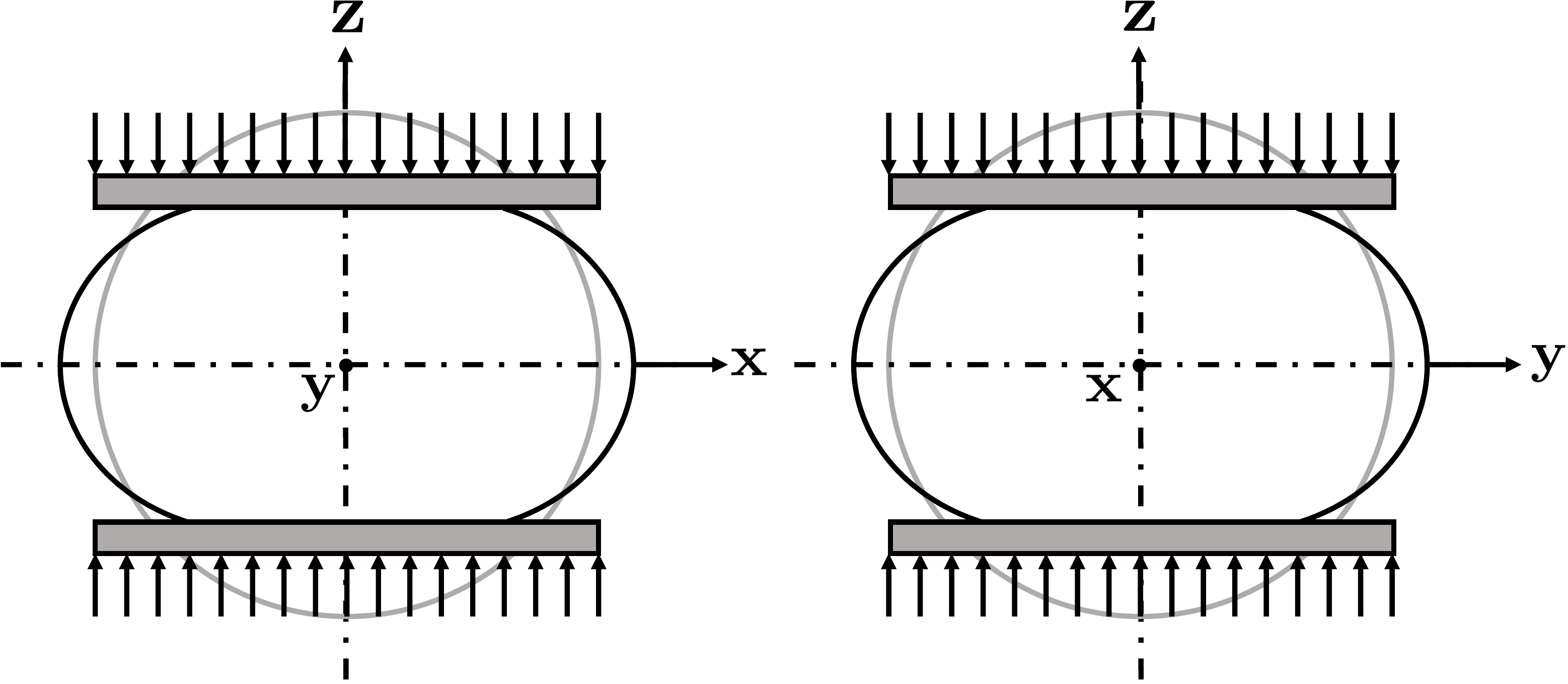}
			\caption{\label{fig:NL_Simple1}}
		\end{subfigure}
		\begin{subfigure}[b]{0.48\linewidth}
			\includegraphics[keepaspectratio,scale=0.27]{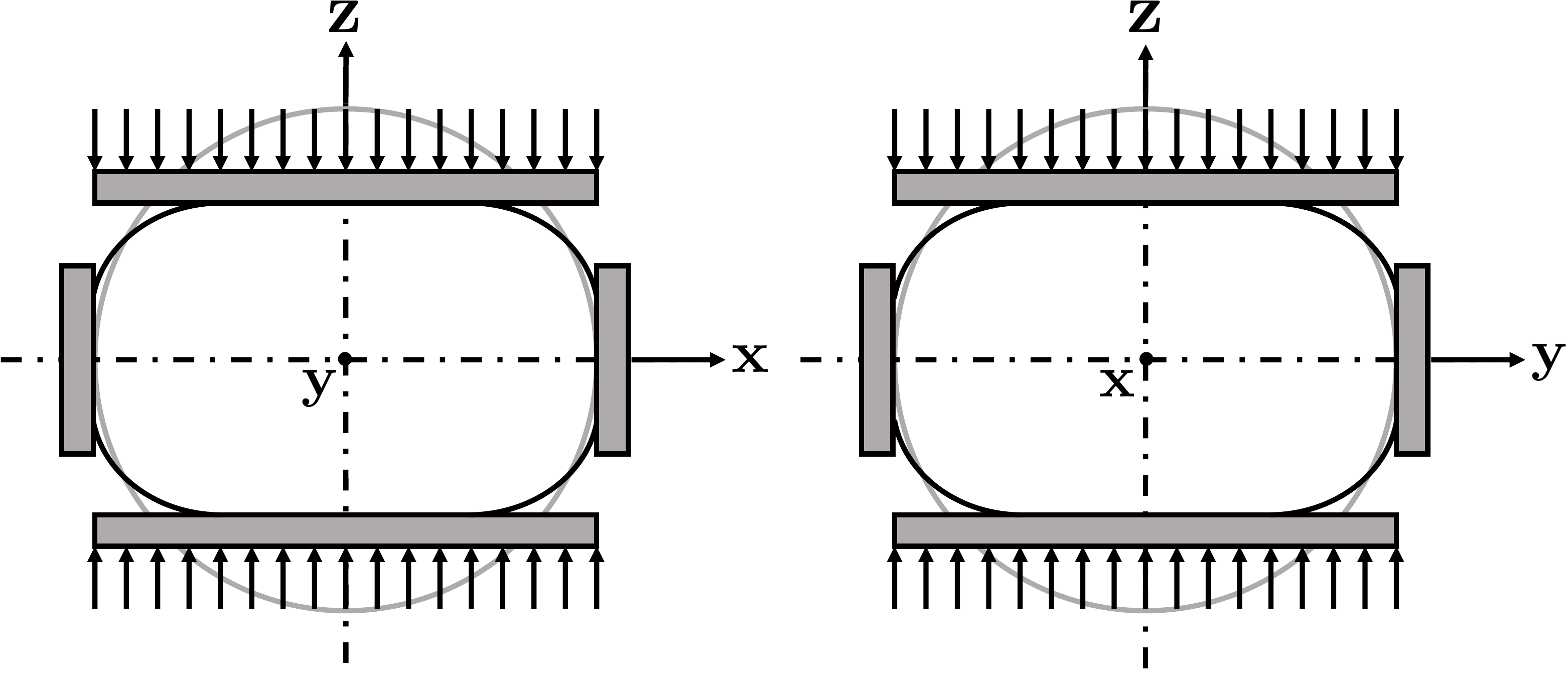}
			\caption{\label{fig:NL_Die1}}
		\end{subfigure}
		\begin{subfigure}[b]{0.48\linewidth}
			\includegraphics[keepaspectratio,scale=0.27]{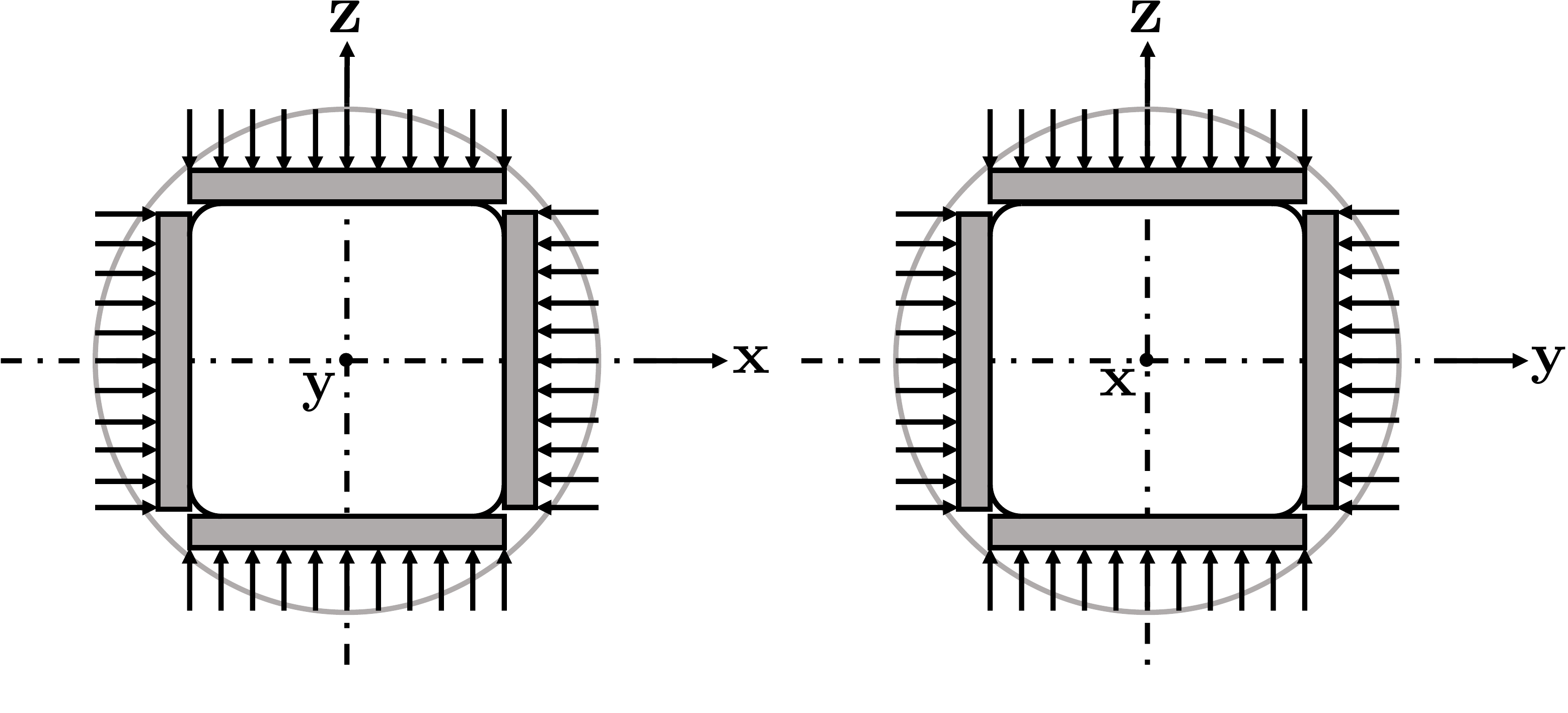}
			\caption{\label{fig:NL_Hydro1}}
		\end{subfigure}
		\begin{subfigure}[b]{0.48\linewidth}
			\includegraphics[keepaspectratio,scale=0.27]{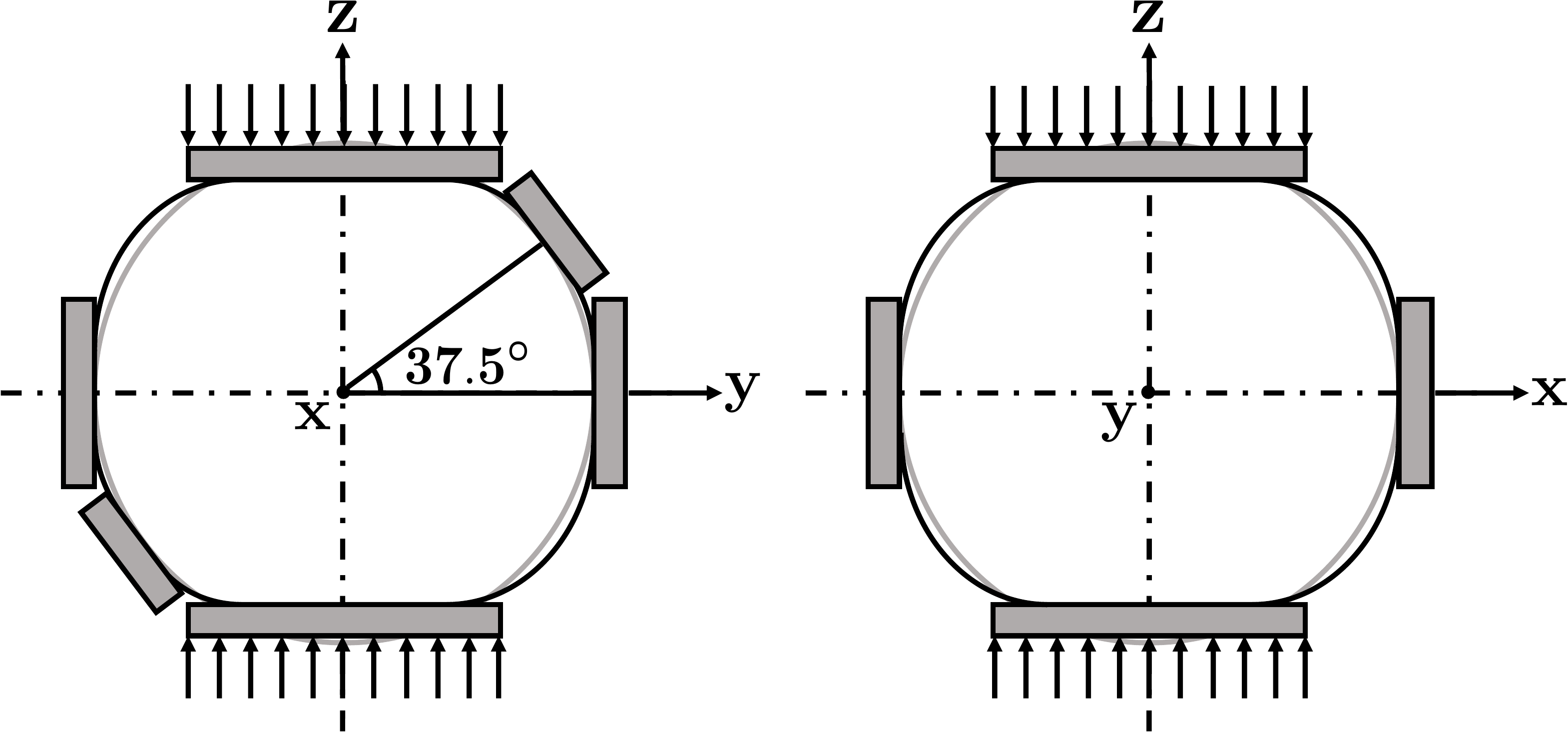}
			\caption{\label{fig:NL_Case41}}
		\end{subfigure}
		\caption{Schematic of the loading configurations considered for validation of contact radius and curvature corrections to the nonlocal contact formulation. (\subref{fig:NL_Simple1}). Simple Compression, (\subref{fig:NL_Die1}). Die Compression, (\subref{fig:NL_Hydro1}). Hydrostatic Compaction and (\subref{fig:NL_Case41}). Die compression with two additional walls at an angle of 37.5$^\circ$ from lateral walls in the y-direction. For each load case, views in x-z and y-z plane are provided for clarity.\label{fig:NL_loadcases}}\bigskip
		\begin{subfigure}[t]{0.76\linewidth}
			\includegraphics[keepaspectratio,scale=0.45]{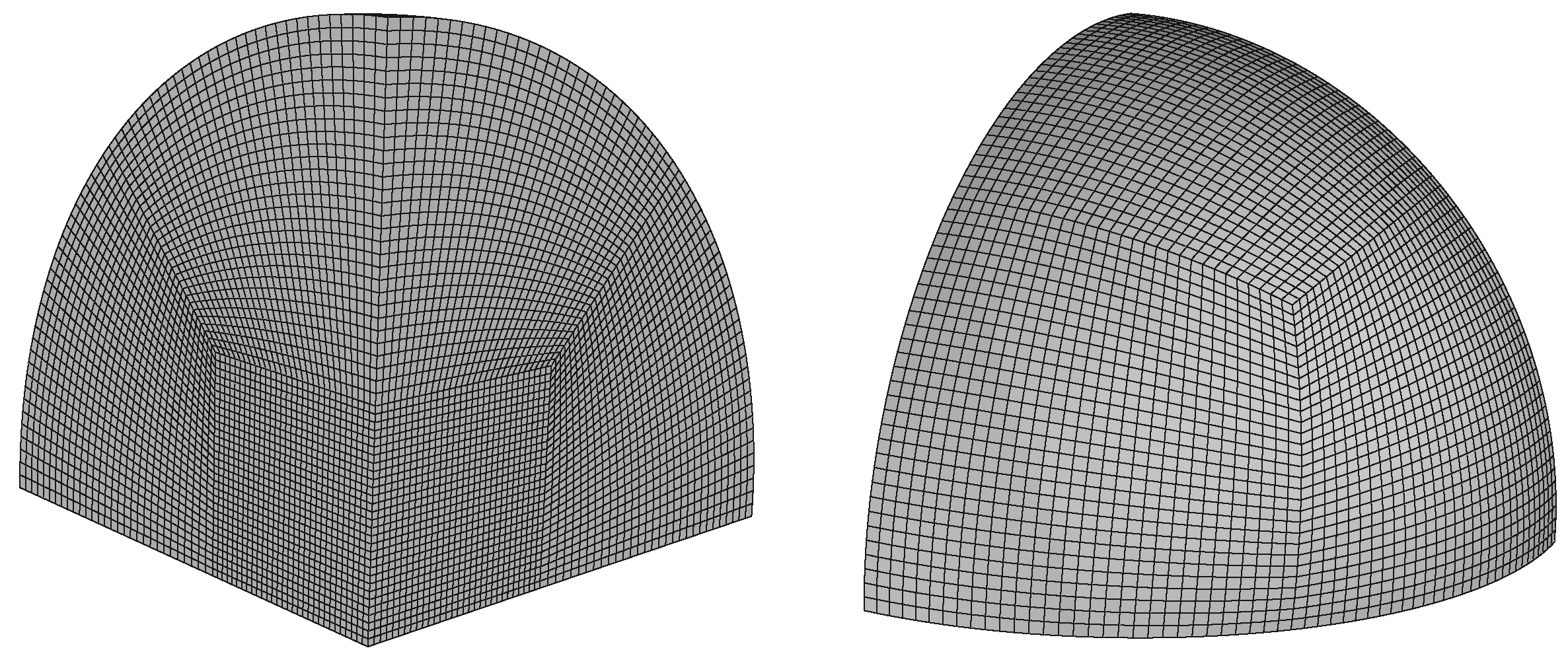}
			\caption{\label{fig:Sphere_mesh_undef}}
		\end{subfigure}
		\begin{subfigure}[t]{0.86\linewidth}
			\includegraphics[keepaspectratio,scale=0.45]{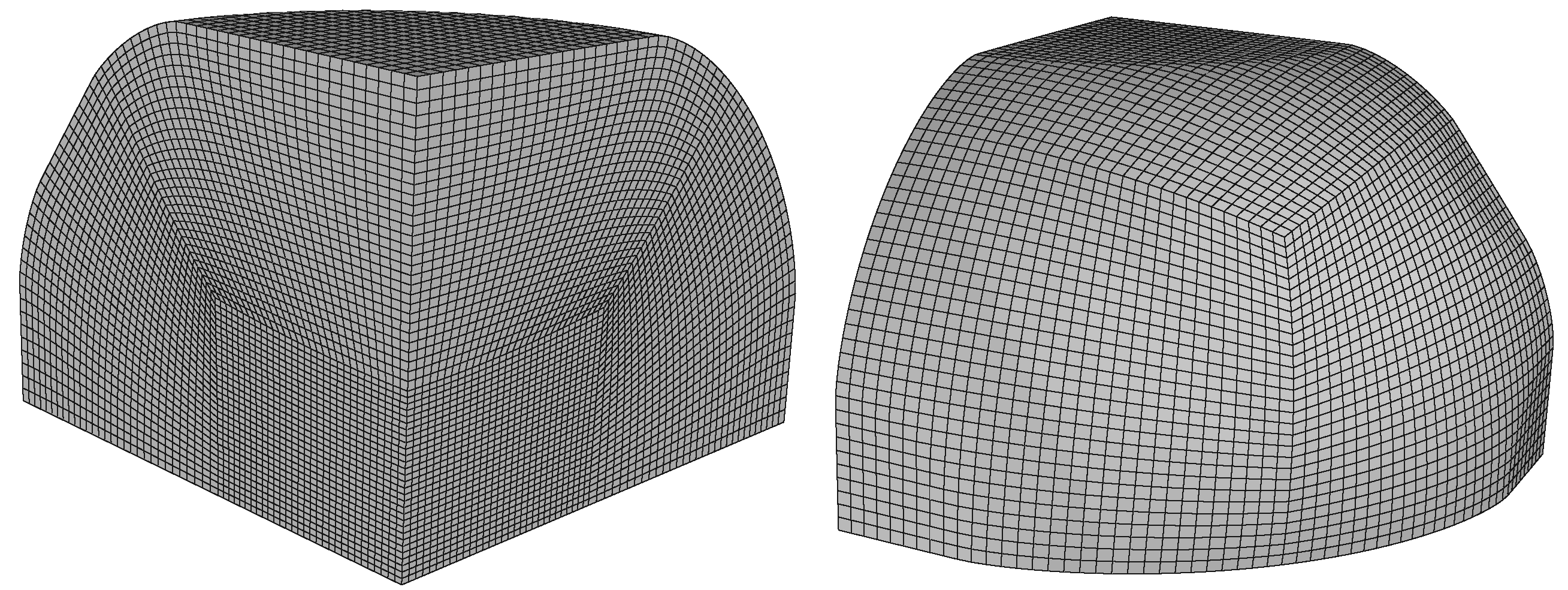}
			\caption{\label{fig:Sphere_mesh_def}}
		\end{subfigure}
		\caption{Finite-element mesh of one-eighth of a sphere created in ABAQUS. The depicted mesh is coarser than the final converged mesh, and consists of 108,000 hexahedral elements and 113,521 nodes. (\subref{fig:Sphere_mesh_undef}). Initial undeformed mesh. (\subref{fig:Sphere_mesh_def}). Deformed mesh for the fourth loading configuration of die compression with two additional oblique walls.\label{fig:Sphere_mesh}}
	\end{figure}

	Validation of contact radius and curvature corrections to the nonlocal contact formulation was performed by considering four types of loading configurations (Figure \ref{fig:NL_loadcases}), three of which are symmetric configurations that simulate the compaction of particles in a simple cubic lattice, namely simple compression (sphere pressed between two rigid plates, figure \ref{fig:NL_Simple1}), die compaction (sphere compressed between rigid plates and constrained laterally by rigid walls, figure \ref{fig:NL_Die1}) and hydrostatic compaction (sphere compressed triaxially by rigid plates, figure \ref{fig:NL_Hydro1}). The fourth configuration is asymmetric,  consisting of two additional walls perpendicular to the y-z plane between the plates and lateral walls of a die compaction configuration (Figure \ref{fig:NL_Case41}). Analytical predictions for all the configurations were compared with detailed finite element simulations, analytical results obtained from the original nonlocal contact formulation \citep{Gonzalez-2012}, and the classical Hertz predictions. For simple compression, predictions were also compared with experimental measurements \citep{Tatara-1989,Tatara-1991b}.
	
	The finite element simulations were performed in ABAQUS on one-eighth of a sphere, owing to geometric and loading symmetries. Finite deformations were considered and the material was characterized as compressible neo-Hookean with an energy density of the form
	\begin{equation} \label{36}
	W(\lambda_1, \lambda_2, \lambda_3) = \frac{\mu}{2}\left[J^{-2/3}(\lambda^2_1+\lambda^2_2+\lambda^2_3)-3\right]+\frac{\lambda}{2}(J-1)^2
	\end{equation}
	where $\lambda_1$, $\lambda_2$ and $\lambda_3$ are the principle material stretches, $J=\det(\mathbf{F})=\lambda_1\lambda_2\lambda_3$ and $\mu$ and $\lambda$ are the Lam\'{e} constants. In ABAQUS, $C_{10} = \mu/2$ and $D_1=2/\lambda$ are the input paramters. The elastic constants used in the simulations correspond to the values of $E=1.85$ MPa and $\nu = 0.48$ reported for rubber \citep{Tatara-1989,Tatara-1991b}. Following a mesh convergence study, the mesh comprising of 500,000 elements of type C3D8R (8-node linear hexahedral) and 515,201 nodes was chosen. For illustration purposes, a course mesh of 108,000 elements is depicted in Figure \ref{fig:Sphere_mesh}.
	
\begin{figure}[t]
		\centering
		\includegraphics[keepaspectratio,scale=0.65]{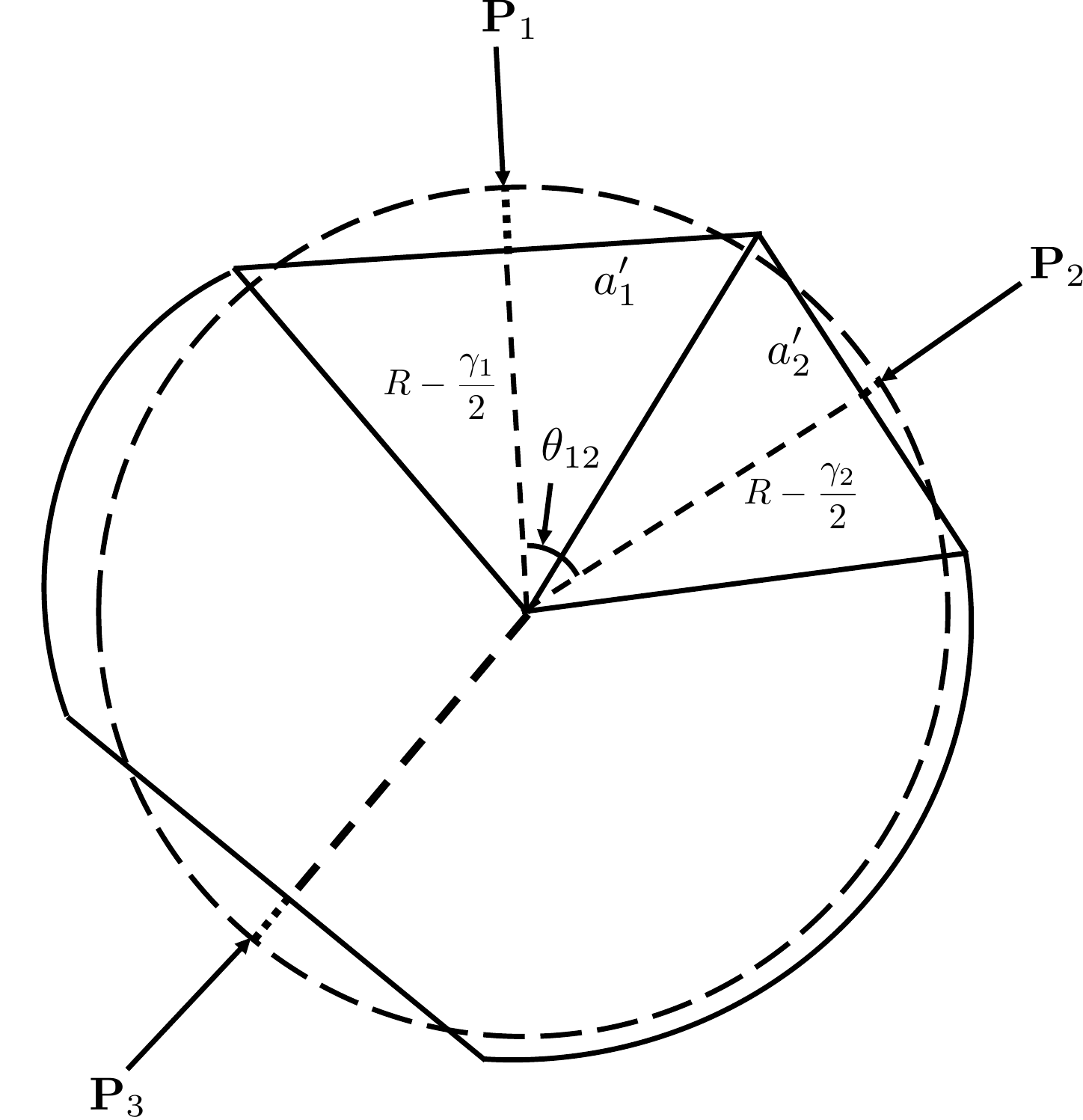}
		\caption{Impingement of two contacts of radii $a'_1$ and $a'_2$ at distances $\left(R-\frac{\gamma_1}{2}\right)$ and $\left(R-\frac{\gamma_2}{2}\right)$ respectively from the center of a particle of radius R, separated by an angular distance $\theta_{12}$.}
		\label{fig:contimp}
	\end{figure}

	One of the most important factors to be considered while comparing analytical predictions with numerical simulations, other than the evolution of contact force and contact radius, is the occurrence of contact impingement. This is a phenomenon that causes the contacts to no longer remain circular and, thereby, restricts the applicability of the nonlocal contact formulation. Figure \ref{fig:contimp} provides a geometrical description of the impingement of two contacts in a spherical particle under a general loading configuration. From the figure, the angular distance $\theta_{12}$ between the two contacts at the inception of impingement is given by
	\begin{equation} \label{37}
	\theta_{12} = \tan^{-1}\left(\frac{a'_1}{R-(\gamma_1/2)}\right)+\tan^{-1}\left(\frac{a'_2}{R-(\gamma_2/2)}\right)
	\end{equation} 

\begin{figure}[!t]
		\centering
		\begin{subfigure}[t]{0.5\linewidth}
			\includegraphics[keepaspectratio,scale=0.68]{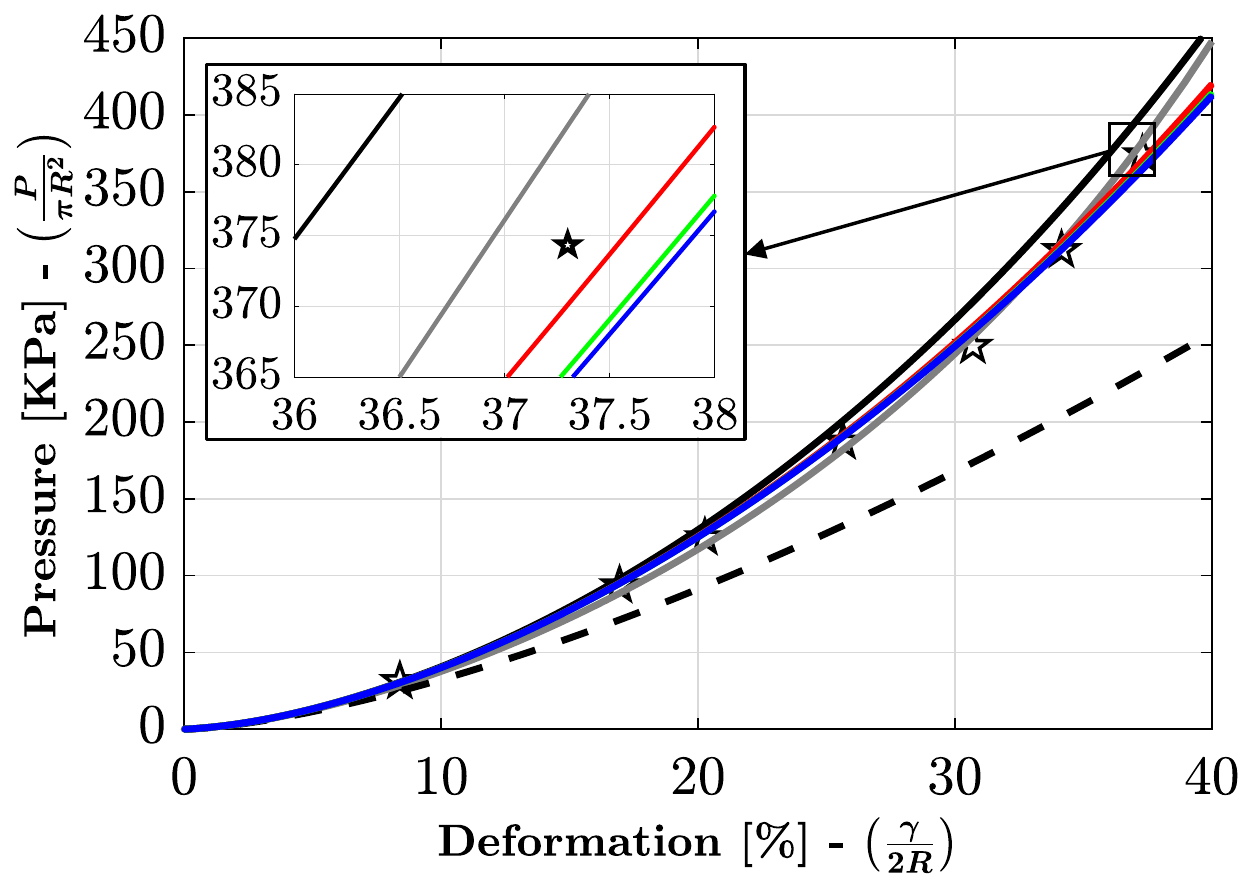}
			\caption{\label{fig:NL_Simple2}}
		\end{subfigure}
		\begin{subfigure}[t]{0.494\linewidth}
			\includegraphics[keepaspectratio,scale=0.68]{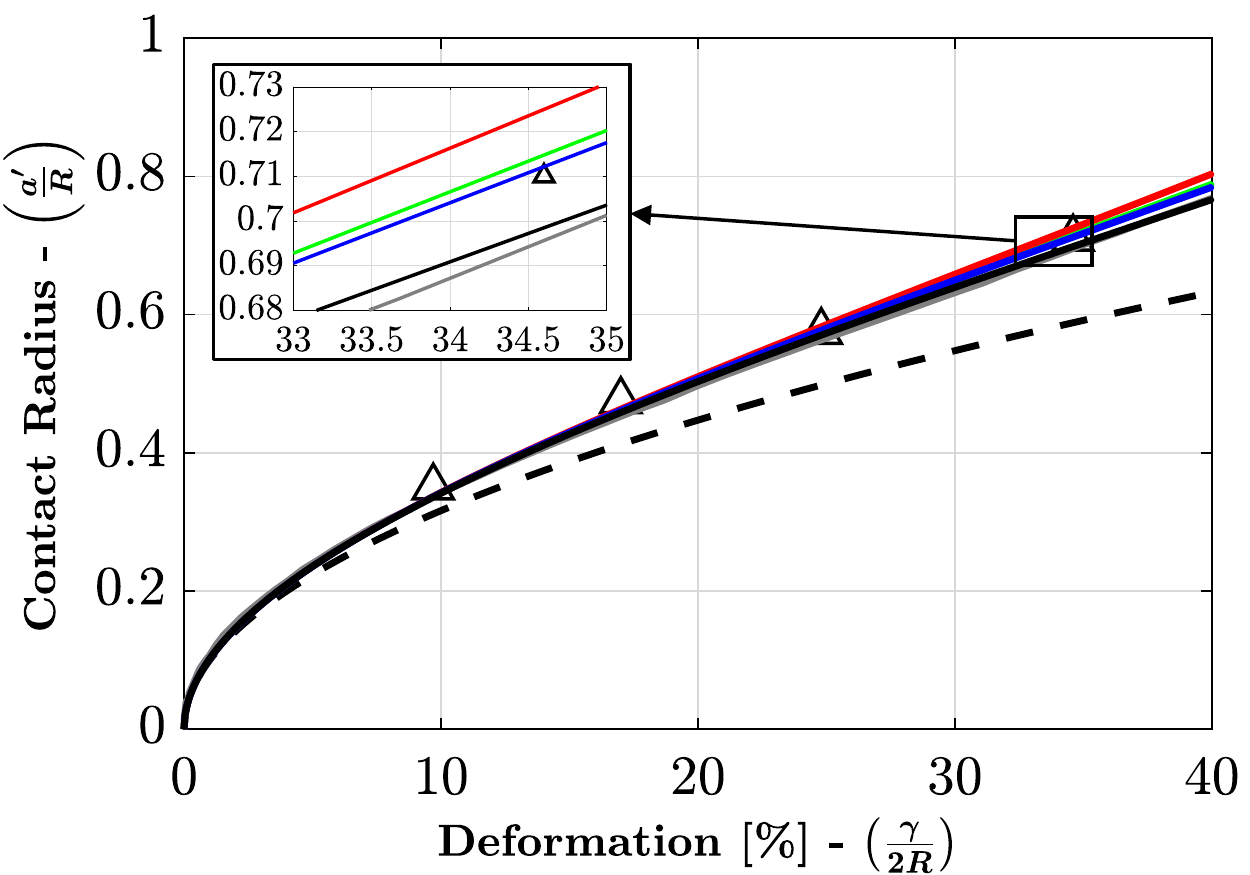}
			\caption{\label{fig:NL_Simple3}}
		\end{subfigure}
		\caption{Load - deformation (\subref{fig:NL_Simple2}) and contact radius - deformation (\subref{fig:NL_Simple3}) curves for simple compression of a rubber sphere. Hertz theory predictions (black-dashed curves), nonlocal contact formulation results without contact radius and curvature corrections (black curves), with contact radius and two- (red curves), three- (green curves) and four- (blue curves) term corrections, finite element solution (grey curves), and experimental measurements \citep{Tatara-1989,Tatara-1991b} (five-pointed stars and triangles) are presented. The convergence of predictions at a four-term correction is shown in the inserts. \label{fig:NL_Simple}}\bigskip 
		\begin{subfigure}[t]{0.5\linewidth}
			\centering
			\includegraphics[keepaspectratio,scale=0.68]{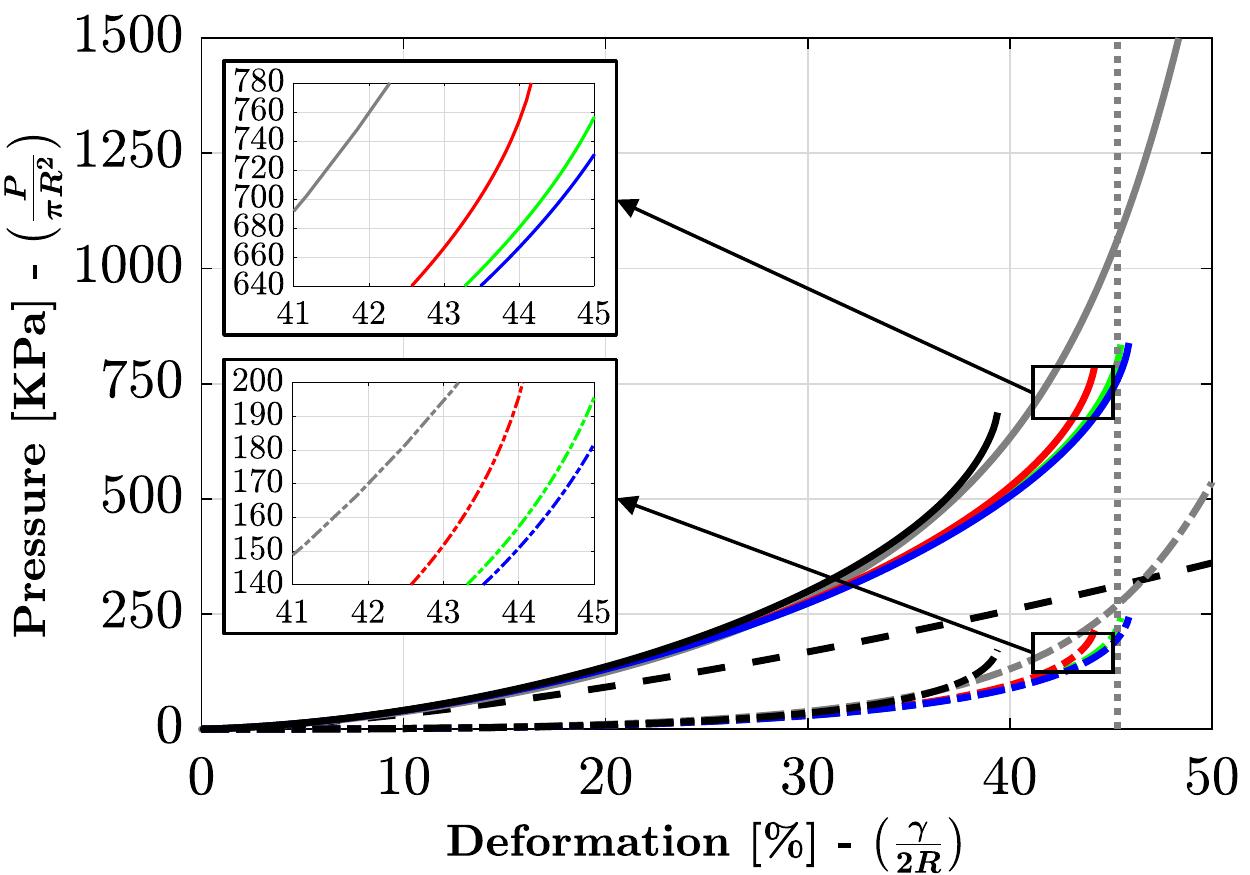}
			\caption{\label{fig:NL_Die2}}
		\end{subfigure}
		\begin{subfigure}[t]{0.494\linewidth}
			\centering
			\includegraphics[keepaspectratio,scale=0.68]{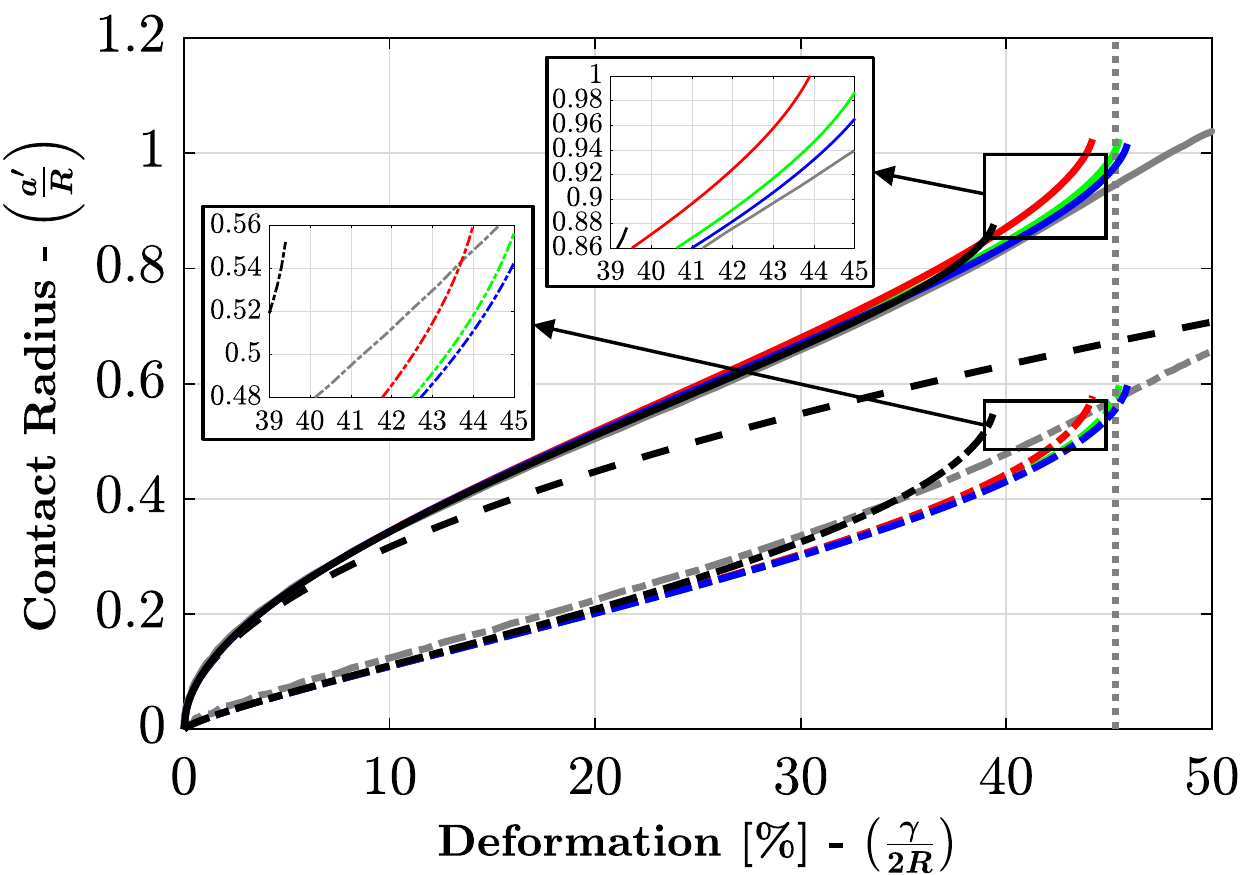}
			\caption{\label{fig:NL_Die3}}
		\end{subfigure}
		\caption{Load - deformation (\subref{fig:NL_Die2}) and contact radius - deformation (\subref{fig:NL_Die3}) curves for die compression of a rubber sphere. Predictions for vertical loaded contacts are given by solid curves while predictions for lateral constrained contacts are given by dashed-dotted curves. Hertz theory predictions (black-dashed curves), nonlocal contact formulation results without contact radius and curvature corrections (black curves), with contact radius and two- (red curves), three- (green curves) and four- (blue curves) term corrections, and finite element solution (grey curves) are presented. The deformation at geometrical contact impingement of vertical and lateral contacts is marked by a grey dotted line. The convergence of predictions at a four-term correction is shown in the inserts.\label{fig:NL_Die}}
	\end{figure}

\begin{figure}[!t]
	\centering
	\begin{subfigure}[t]{0.5\linewidth}
		\centering
		\includegraphics[keepaspectratio,scale=0.68]{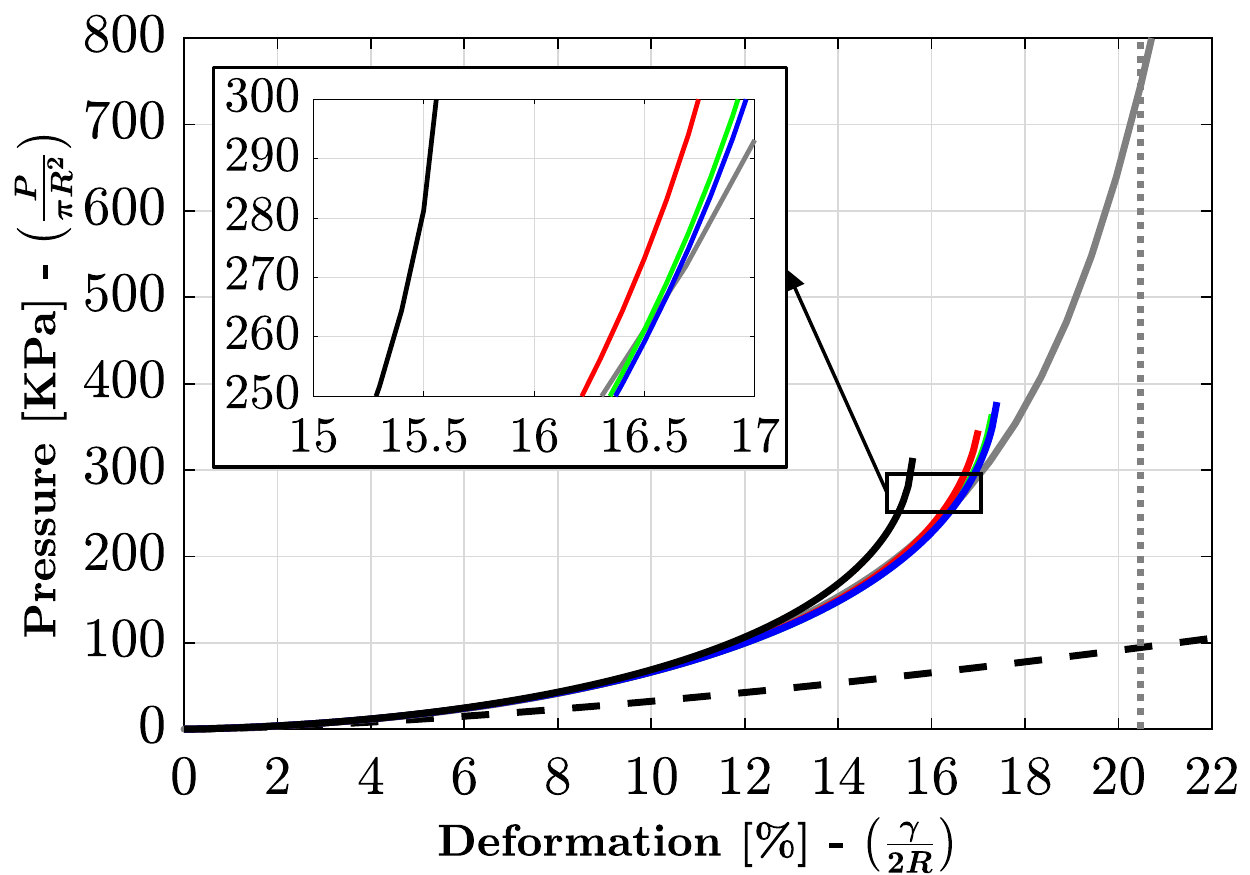}
		\caption{\label{fig:NL_Hydro2}}
	\end{subfigure}
	\begin{subfigure}[t]{0.494\linewidth}
		\centering
		\includegraphics[keepaspectratio,scale=0.68]{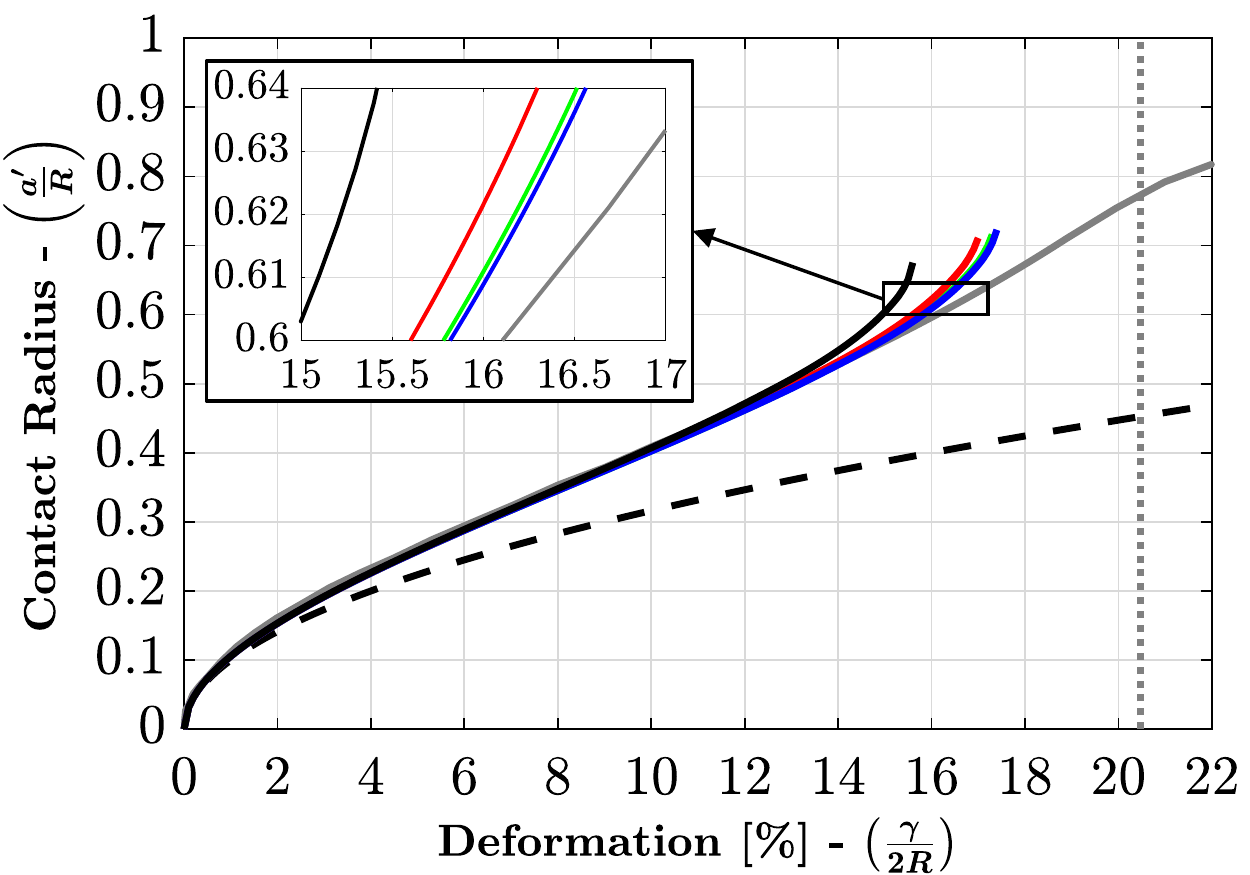}
		\caption{\label{fig:NL_Hydro3}}
	\end{subfigure}
	\caption{Load - deformation (\subref{fig:NL_Hydro2}) and contact radius - deformation (\subref{fig:NL_Hydro3}) curves for hydrostatic compression of a rubber sphere. Hertz theory predictions (black-dashed curves), nonlocal contact formulation results without contact radius and curvature corrections (black curves), with contact radius and two- (red curves), three- (green curves) and four- (blue curves) term corrections, and finite element solution (grey curves)) are presented. The deformation at geometrical contact impingement of contacts is marked by a grey dotted line. The convergence of predictions at a four-term correction is shown in the inserts. \label{fig:NL_Hydro}}\bigskip
	\begin{subfigure}[t]{0.5\linewidth}
		\centering
		\includegraphics[keepaspectratio,scale=0.68]{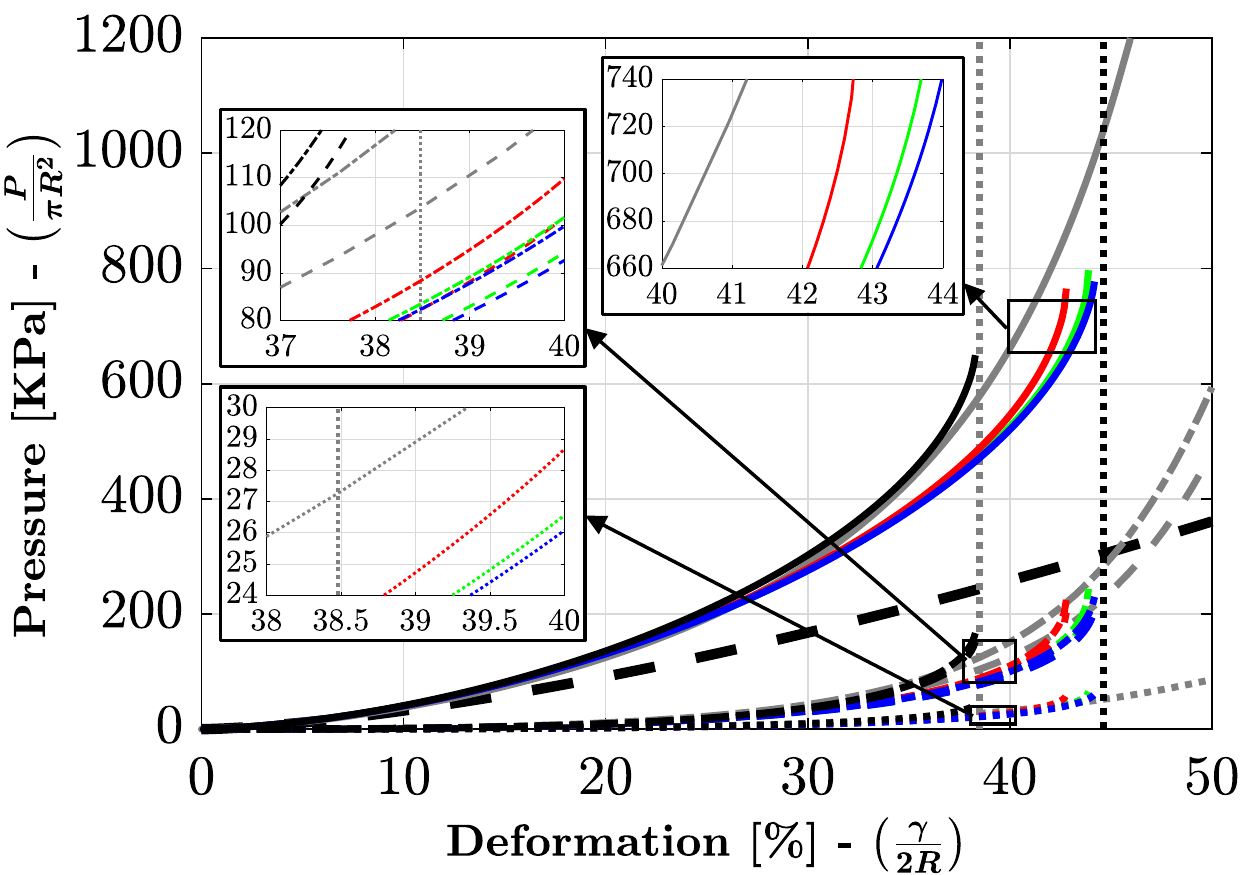}
		\caption{\label{fig:NL_Case42}}
	\end{subfigure}
	\begin{subfigure}[t]{0.494\linewidth}
		\centering
		\includegraphics[keepaspectratio,scale=0.68]{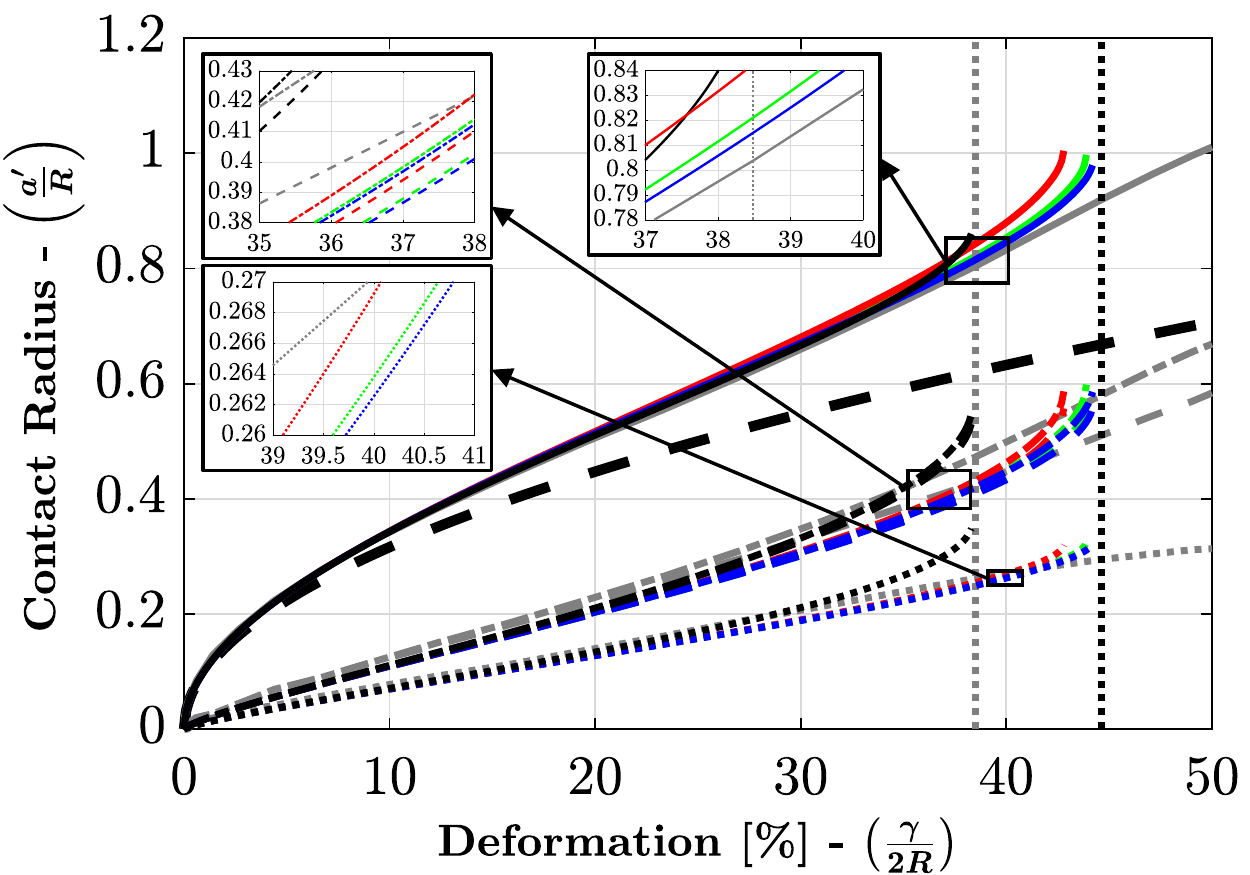}
		\caption{\label{fig:NL_Case43}}
	\end{subfigure}
	\caption{Load - deformation (\subref{fig:NL_Case42}) and contact radius - deformation (\subref{fig:NL_Case43}) curves for a rubber sphere under an asymmetric loading configuration of die compression with oblique walls. Predictions are represented by solid curves for vertical loaded contacts, dashed-dotted curves for lateral constrained contacts in x-direction, dashed curves for lateral constrained contacts in y-direction, and dotted curves for oblique contacts. Hertz theory predictions (bold black-dashed curves), nonlocal contact formulation results without contact radius and curvature corrections (black curves), with contact radius and two- (red curves), three- (green curves) and four- (blue curves) term corrections, and finite element solution (grey curves) are presented. The deformation at geometrical contact impingement of vertical, oblique and lateral (y-direction) contacts is marked by a grey dotted line, and that of vertical contacts and lateral contacts in x-direction by a black dotted line. The convergence of predictions at a four-term correction is shown in the inserts.\label{fig:NL_Case4}}
	\end{figure}

	Figures \ref{fig:NL_Simple}-\ref{fig:NL_Case4} present the validation results in form of load-versus-deformation ($F/\pi R^2$ versus $\gamma/2R$) and contact radius-versus-deformation ($a'/R$ versus $\gamma/2R$) plots. The figures also show the values of deformation at which contact impingement occurs for each finite-element simulation (except simple compression where impingement does not occur)---this information is extracted by using the contact radius and displacement values obtained from the simulations in eq. \ref{37}. It is interesting to note that for all loading configurations except simple compression, the analytical predictions show an apparent strain-hardening or stiffening at various deformation levels. This is an artifact of the formulation, which possibly results from various modeling assumptions made for evaluation of $\gamma^\mathrm{NL}$ \citep{Gonzalez-2012}, that limits the range of applicability of the nonlocal contact formulation. Such limited range of applicability exists for simple compression as well, albeit at deformations greater than 80\% that are beyond the scope of this study.
	
	Analysis of the plots suggests that the proposed contact radius and curvature corrections increase the range of applicability of the nonlocal contact formulation, thereby enabling predictions at higher levels of deformation. Further improvement is observed with increase in the order of curvature correction until convergence, which is achieved with a four-term correction. Quantitatively, the range of applicability of the formulation is increased by roughly 5\% for die compaction, 2\% for hydrostatic compaction and 6\% for die compaction with oblique walls. Additionally, the corrections enable predictions closer to the geometric contact impingement, marked by dotted straight lines in the graphs. This is well represented in the graphs for die compaction, where predictions continue right until the impingement of vertical and lateral contacts, and for die compaction with additional walls, where predictions continue until the impingement of vertical and lateral contacts in the x-direction. Interestingly, analytical predictions for the fourth configuration are remarkably accurate even after the impingement of vertical, oblique and lateral (y-direction) contacts (grey dotted line), indicating that predictions of the extended nonlocal contact formulation are accurate until impingement of all particle contacts. As for the case of simple compression, an overall improvement in the representation of experimental measurements for both contact force and radius is observed, which again converges at the four-term curvature correction.
	
	\section{Summary and discussion} \label{sec6}
	
	We have developed the analytical framework for contact radius correction to the nonlocal contact formulation, which accounts for local and nonlocal contributions to the radial deformation of contact boundaries due to multiple contact forces acting on a single particle. Furthermore, we have provided a method of curvature correction to relax the traditional assumption of one-term Taylor series representation of undeformed contacting surfaces.  For definiteness, we have restricted attention to elastic spheres in the absence of gravitational forces, adhesion or friction. Hence, a notable feature of the contact formulation presented here is that, when no contact radius and curvature corrections are accounted for, it reduces to the nonlocal contact formulation presented by \cite{Gonzalez-2012} and thus, when all nonlocal effects and corrections are neglected, it reduces to Hertz theory. Another salient feature of the proposed formulation is that it increases the range of applicability of the nonlocal contact formulation \citep*{Gonzalez-2012}; consequently, it enables accurate predictions of contact behavior until contact impingement for confined loading configurations. Specifically, we have investigated four different loading conditions (namely simple compression, die compression within four walls and within six walls, and hydrostatic compaction) and have successfully compared the predictions of the proposed nonlocal formulation with experimental and detailed finite-element simulation results of rubber particles.

	We close by pointing out future research directions and possible approaches for extension of the formulation.
	
	First, the work presented in this paper together with the nonlocal contact formulation by \cite{Gonzalez-2012} serve as the foundation for conceiving an analytical elasto-plastic nonlocal contact formulation. Like elastic particles, the assumption of independent contacts is not valid for particles deforming predominantly plastically in the range of moderate to high mesoscopic deformations \citep*{Mesarovick-2000}. Recenty, finite element simulations of linear elastic-perfectly plastic particles \citep*{Tsigginos-2015} have shown contact interactions at moderate deformations physically manifesting as coalescence of local plastically deforming zones around individual contacts, with further deformation leading to the development of a low-compressibility regime when the contact deformations become predominantly elastic. The systematic investigation of these interaction effects is a worthwhile direction of future research.
	
	Second, although the work presented in this paper is a contribution to predictive modeling of confined granular systems, considerable effort is required for modeling of contact behavior beyond contact impingement. Finite element simulations presented in this paper show an almost linear increase in contact force with deformation after impingement of all contacts, indicating a linear force-displacement relationship in this deformation regime. However, rigorous analysis of this linear correlation is desirable, if beyond the scope of this paper.

	\section*{Acknowledgments}
	The authors gratefully acknowledge the support received from the National Science Foundation grant number CMMI-1538861, from Purdue University's startup funds, and from the Network for Computational Nanotechnology (NCN) and nanoHUB.org. Ankit Agarwal also acknowledges the Frederick N. Andrews Fellowship from Purdue University.
	
	\appendix
	
	\section{Determination of radial displacement of a contact boundary point due to a single force} \label{appA}
	
\begin{figure}[t]
		\centering
		\includegraphics[keepaspectratio,scale=0.65]{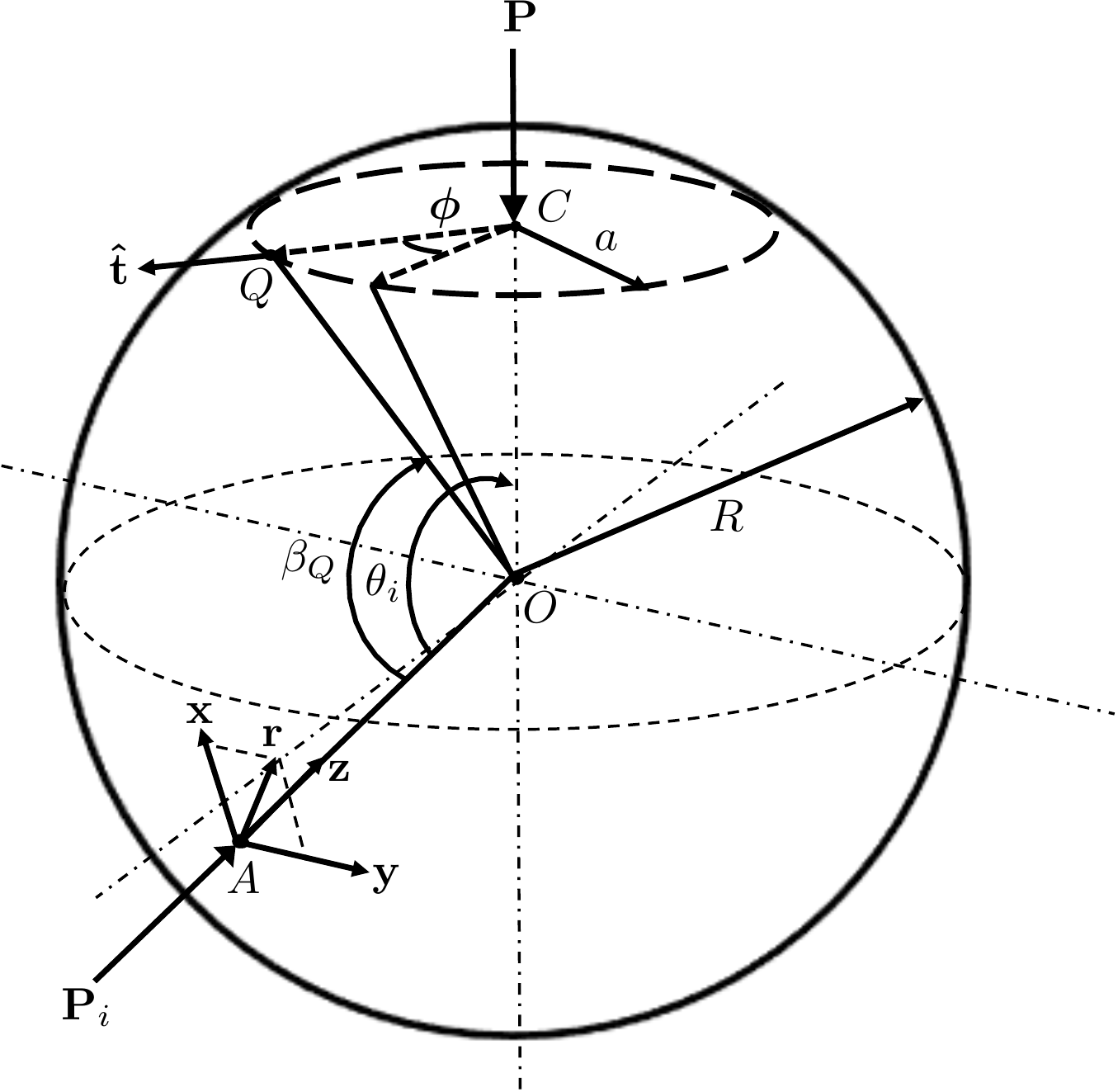}
		\caption{A linear-elastic sphere of radius $R$ under the action of an ellipsodially distributed pressure, approximated by an effective force $P$ on a spherical cap of radius $a$ and center $C$, with concentrated force $P_i$ acting normally on one of its surface points.}
		\label{fig:thetavary}
	\end{figure}

	Figure \ref{fig:thetavary} depicts a three dimensional view of a linear-elastic spherical particle under the action of a concentrated force $P_i$ applied at the origin $A$ of cylindrical coordinates $(z, r)$. We consider the deformation of a spherical cap of base radius $a$ and center $C$, situated at an angular distance $\theta_i$ from the force $P_i$, due to an ellipsoidally distributed pressure given by eq. \ref{1} in section \ref{sec2}. The pressure distribution is approximated by an effective force $P$ applied at $C$. A point $Q$ on the cap boundary is situated at angle $\phi$ from the plane defined by points $A$, $O$ and $C$. Using vector algebra, the angle $AOQ$, denoted by $\beta_Q$, can be expressed as
	\begin{equation} \label{39}
	\beta_Q = \cos^{-1}\left[\cos \left|\theta_i - \sin^{-1}\left(\frac{a}{R}\right)\right| - \left(\frac{a}{R}\right)(1-\cos \phi)\sin\theta_i\right]
	\end{equation}
	The above equation implies that the angular distance between a contact force and any point on the cap boundary can be represented in terms of a constant reference angle $\theta_i$ and a variable angle $\phi$. The dependency on angle $\phi$ vanishes for the two extreme values of $\theta_i$, i.e. $0$ and $\pi$. For $\theta_i=0$, which corresponds to the case when $P_i=P$, eq. \ref{39} reduces to
	\begin{equation} \label{40}
	\left.\beta_Q\right|_{\theta_i=0} =  \sin^{-1}\left(\frac{a}{R}\right)
	\end{equation}
	the above result restates the geometrical fact that the contact force $P$ is equidistant from any point on the cap boundary, the angular distance being $\sin^{-1}(a/R)$. A similar result is obtained for $\theta_i=\pi$, which makes the force $P_i$ equidistant from any point on the cap boundary. The angular distance in this case becomes
	\begin{equation} \label{41}
	\left.\beta_Q\right|_{\theta_i=\pi} =\pi- \sin^{-1}\left(\frac{a}{R}\right)
	\end{equation}
	
	At point $Q$ with coordinates $(z, r): \left(2R\sin^2(\beta_Q/2), R\sin\beta_Q\right)$, displacements due to force $P_i$ along the $(z, r)$ axes, denoted by  $ (W_{i,Q}, U_{i,Q})$, are given by the Boussinesq solution \citep*{Johnson-1985,Timo-1970} and can be expressed in terms of angle $\beta_Q$ as
	\begin{equation} \label{42}
	W_{i,Q}(\beta_Q) =\frac{(1+\nu){P_i}}{2\pi ER}\left[\frac{\sin (\beta_Q/2)}{2}+\frac{1-\nu}{\sin(\beta_Q/2)}\right]
	\end{equation}
	\begin{equation} \label{43}
	U_{i,Q}(\beta_Q) = \frac{(1+\nu){P_i}}{4\pi ER}\left[\cos(\beta_Q/2)-\frac{(1-2\nu)\cos(\beta_Q/2)}{\sin(\beta_Q/2)\left(1+\sin(\beta_Q/2)\right)}\right]
	\end{equation} 
	
	Similarly, for the cap center $C$ with coordinates $(z, r): \left(2R\sin^2(\theta_i/2), R\sin\theta_i\right)$, displacements due to force $P_i$ along the $(z, r)$ axes, denoted by $(W_{i,C}, U_{i,C})$, can be expressed in terms of angle $\theta_i$ as
	\begin{equation} \label{44}
	W_{i,C}(\theta_i) =\frac{(1+\nu){P_i}}{2\pi ER}\left[\frac{\sin (\theta_i/2)}{2}+\frac{1-\nu}{\sin(\theta_i/2)}\right]
	\end{equation}
	\begin{equation} \label{45}
	U_{i,C}(\theta_i) = \frac{(1+\nu){P_i}}{4\pi ER}\left[\cos(\theta_i/2)-\frac{(1-2\nu)\cos(\theta_i/2)}{\sin(\theta_i/2)\left(1+\sin(\theta_i/2)\right)}\right]
	\end{equation}
	The displacement of point $Q$ due to force $P_i$ in the radial direction with respect to center $C$, denoted by $u_{i,Q}$, can be expressed as
	\begin{equation} \label{46}
	u_{i,Q} = \left[(W_{i,Q} - W_{i,C})\hat{e}_z+(U_{i,Q}-U_{i,C})\hat{e}_{r}\right]\cdot\hat{t}
	\end{equation} 
	where $\hat{t}$ is a unit vector in the radial direction at $Q$, as shown in figure \ref{fig:thetavary}.  Using vector algebra, it can be proved that the vector $\hat{t}$ can be expressed in terms of $(x, y, z)$ coordinates as
	\begin{equation} \label{47}
	\hat{t} = -\cos\theta_i\cos\phi\,\hat{e}_x - \sin\phi\,\hat{e}_y-\sin\theta_i\cos\phi\,\hat{e}_z
	\end{equation}
	which implies that $\hat{t}$ in terms of $(z, r)$ coordinates becomes
	\begin{equation} \label{48}
	\hat{t} = -\sin\theta_i\cos\phi\hat{e}_z + \left(\sqrt{1-\sin^2\theta_i\cos^2\phi}\right)\hat{e}_r
	\end{equation}
	Substituting $\hat{t}$ from eq. \ref{48} into eq. \ref{46}, we get
	\begin{equation} \label{49}
	u_{i,Q} = -(W_{i,Q} - W_{i,C})\sin\theta_i\cos\phi+(U_{i,Q}-U_{i,C})\sqrt{1-\sin^2\theta_i\cos^2\phi}
	\end{equation}
	
	Eq. \ref{49} provides a general definition of the radial displacement of a contact boundary point on an elastic spherical particle due to a concentrated force acting on the surface of the particle. After substituting the values of $W_{i,Q}$, $U_{i,Q}$, $W_{i,C}$ and $U_{i,C}$ from eqs. \ref{42}, \ref{43}, \ref{44} and \ref{45} respectively into eq. \ref{49}, the resulting final expression for $u_{i,Q}$ in terms of angles $\theta_i$ and $\phi$ is given by
	\begin{equation} \label{50}
	\begin{aligned}
	u_{i,Q} &=\frac{(1+\nu){P_i}}{2\pi ER}\left[\frac{\sin\theta_i\cos\phi\left(\sin(\theta_i/2)-\sin(\beta_Q(\theta_i,\phi)/2)\right)\left(\sin(\theta_i/2)\sin(\beta_Q(\theta_i,\phi)/2)-2+2\nu\right)}{2\sin(\theta_i/2)\sin(\beta_Q(\theta_i,\phi)/2)}\right. \\
	&\quad\left.+\sqrt{1-\sin^2\theta_i\cos^2\phi}\left[\frac{1}{2}\left(\cos(\beta_Q(\theta_i,\phi)/2)-\cos(\theta_i/2)\right)\right.\right. \\
	&\quad\left.\left.-(1-2\nu)\left(\frac{1-\sin(\beta_Q(\theta_i,\phi)/2)}{\sin\beta_Q(\theta_i,\phi)}-\frac{\cos(\theta_i/2)}{2\sin(\theta_i/2)(1+\sin(\theta_i/2))}\right)\right]\vphantom{\frac{\sin\theta_i\cos\phi\left(\sin\frac{\theta_i}{2}-\sin(\beta_Q(\theta_i,\phi)/2)\right)\left(\frac{1}{2}\sin(\theta_i/2)\sin(\beta_Q(\theta_i,\phi)/2)-1+\nu\right)}{\sin(\theta_i/2)\sin(\beta_Q(\theta_i,\phi)/2)}}\right]
	\end{aligned}
	\end{equation}
	The above expression can be further reduced to the following form
	\begin{equation} \label{51}
	u_{i,Q} = \frac{P_i}{\eta_{i,Q}}
	\end{equation}
	where,
	\begin{equation} \label{52}
	\begin{aligned}
	\frac{1}{\eta_{i,Q}} &= \frac{1+\nu}{2\pi ER}\left[\frac{\sin\theta_i\cos\phi\left(\sin(\theta_i/2)-\sin(\beta_Q(\theta_i,\phi)/2)\right)\left(\sin(\theta_i/2)\sin(\beta_Q(\theta_i,\phi)/2)-2+2\nu\right)}{2\sin(\theta_i/2)\sin(\beta_Q(\theta_i,\phi)/2)}\right. \\
	&\quad\left.+\sqrt{1-\sin^2\theta_i\cos^2\phi}\left[\frac{1}{2}\left(\cos(\beta_Q(\theta_i,\phi)/2)-\cos(\theta_i/2)\right)\right.\right. \\
	&\quad\left.\left.-(1-2\nu)\left(\frac{1-\sin(\beta_Q(\theta_i,\phi)/2)}{\sin\beta_Q(\theta_i,\phi)}-\frac{\cos(\theta_i/2)}{2\sin(\theta_i/2)(1+\sin(\theta_i/2))}\right)\right]\vphantom{\frac{\sin\theta_i\cos\phi\left(\sin\frac{\theta_i}{2}-\sin(\beta_Q(\theta_i,\phi)/2)\right)\left(\frac{1}{2}\sin(\theta_i/2)\sin(\beta_Q(\theta_i,\phi)/2)-1+\nu\right)}{\sin(\theta_i/2)\sin(\beta_Q(\theta_i,\phi)/2)}}\right]
	\end{aligned}
	\end{equation}
	
	\section{Calculation of pressure distribution on the contact surface of an elastic sphere} \label{appB}
	
	Considering the contact configuration presented in Figure \ref{fig:NL_contact} in Section \ref{sec3}, the contact area is circular and depicted in Figure \ref{fig:pressureprofile} in Section \ref{sec4}. Based on the analysis of \cite{LUO-1958} and \cite{Cattaneo-1947}, the following approximate form of pressure distribution can be assumed at a radial distance $r_{ij}$.
	\begin{equation} \label{53}
	p_{ij}(r_{ij}) = \sum_{n = 1}^{N}\rho_n\left(1-\frac{r^2_{ij}}{a^2_{ij}}\right)^{\frac{2n-1}{2}}
	\end{equation}
	where $N$ corresponds to the number of Taylor series terms considered in the profile function, and $\rho_n$ are $N$ unknown function parameters. For an internal point $A$ in the contact region at a radial distance $r_{ij}$ from the contact center, we can write the following using cosine rule
	\begin{equation} \label{54}
	s^2_{ij} = r^2_{ij} + q^2_{ij} + 2r_{ij}q_{ij}\cos{}\omega_{ij}
	\end{equation} 
	Using eqs. \ref{53} and \ref{54}, the pressure distribution at elemental region $B(q_{ij}, \omega_{ij})$ can be written as
	\begin{equation} \label{55}
	p_{ij}(q_{ij}, \omega_{ij}) =\sum_{n = 1}^{N}\frac{\rho_n}{a^{2n-1}_{ij}}(a^2_{ij} - r^2_{ij}-2r_{ij}q_{ij}\cos{}\omega_{ij} - q^2_{ij})^{\frac{2n-1}{2}} 
	\end{equation}
	Let $\zeta_{ij} = \sqrt{a^2_{ij}-r^2_{ij}}$ and $\delta_{ij} = r_{ij}\cos{}\omega_{ij}$. Then eq. \ref{54} reduces to
	\begin{equation} \label{56}
	p_{ij}(q_{ij}, \omega_{ij}) = \sum_{n = 1}^{N}\frac{\rho_n}{a^{2n-1}_{ij}}(\zeta^2_{ij}-2\delta_{ij}q_{ij} - q^2_{ij})^{\frac{2n-1}{2}}
	\end{equation}
	Substituting the expression for $p_{ij}(q_{ij}, \omega_{ij})$ given by eq. \ref{56} in eq. \ref{22} (Section \ref{sec4}), the displacement field inside the circular region now becomes
	\begin{equation} \label{57}
	\begin{aligned}
	w_i(r_{ij})+w_j(r_{ij})&=\bar{u}_{iz_{ij}}(r_{ij})+\bar{u}_{jz_{ij}}(r_{ij}) \\
	&=\left(\frac{1-\nu^2_i}{\pi{E_i}}+\frac{1-\nu^2_j}{\pi{E_j}}\right)\int_{0}^{2\pi}d\omega_{ij}\int_{0}^{q_1}\left\{\sum_{n = 1}^{N}\frac{\rho_n}{a^{2n-1}_{ij}}(\zeta^2_{ij}-2\delta_{ij}q_{ij} - q^2_{ij})^{\frac{2n-1}{2}}\right\}dq_{ij}
	\end{aligned}
	\end{equation}
	where $q_1$ is the positive root of the equation \cite[pg. 60]{Johnson-1985} 
	\begin{equation} \label{58}
	q^2_{ij} + 2\delta_{ij}q_{ij} - \zeta^2_{ij}  = 0
	\end{equation}
	given by
	\begin{equation} \label{59}
	q_1 = -\delta_{ij}  + \sqrt{\zeta^2_{ij}  + \delta^2_{ij} }
	\end{equation}
	We now consider three cases depending upon the number of Taylor series terms ($N$) taken in the profile function. The three cases correspond to $N = 2$, $3$ and $4$. We shall calculate the pressure distribution for each of these cases.
	
	We first calculate the displacement field for $N=4$ and then modify the function accordingly for different cases. First, the internal integrals with respect to $dq_{ij}$ for $n=1$, $2$, $3$ and $4$ are obtained and given by  
	\begin{equation} \label{60}
	\int_{0}^{q_1}(\zeta^2_{ij}-2\delta_{ij}q_{ij} - q^2_{ij})^{1/2}dq_{ij} = -\frac{1}{2}\zeta_{ij}\delta_{ij} + \frac{1}{2}(\zeta^2_{ij} + \delta^2_{ij})\left\{\frac{\pi}{2} - \tan^{-1}\left(\frac{\delta_{ij}}{\zeta_{ij}}\right)\right\}
	\end{equation}
	\begin{equation} \label{61}
	\int_{0}^{q_1}(\zeta^2_{ij}-2\delta_{ij}q_{ij} - q^2_{ij})^{3/2}dq_{ij}  =-\frac{1}{8}\zeta_{ij}\delta_{ij}(5\zeta^2_{ij}+3\delta^2_{ij}) + \frac{3}{8}(\zeta^2_{ij} + \delta^2_{ij})^2\left\{\frac{\pi}{2} - \tan^{-1}\left(\frac{\delta_{ij}}{\zeta_{ij}}\right)\right\}
	\end{equation}
	\begin{equation} \label{62}
	\int_{0}^{q_1}(\zeta^2_{ij}-2\delta_{ij}q_{ij} - q^2_{ij})^{5/2}dq_{ij} = -\frac{1}{48}\zeta_{ij}\delta_{ij}(33\zeta^4_{ij}+40\zeta^2_{ij}\delta^2_{ij}+15\delta^4_{ij}) + \frac{5}{16}(\zeta^2_{ij} + \delta^2_{ij})^3\left\{\frac{\pi}{2} - \tan^{-1}\left(\frac{\delta_{ij}}{\zeta_{ij}}\right)\right\}
	\end{equation}
	\begin{equation} \label{63}
	\begin{aligned}
	\int_{0}^{q_1}(\zeta^2_{ij}-2\delta_{ij}q_{ij} - q^2_{ij})^{7/2}dq_{ij} &= -\frac{1}{384}\zeta_{ij}\delta_{ij}(279\zeta^6_{ij}+511\zeta^4_{ij}\delta^2_{ij}+385\zeta^2_{ij}\delta^4_{ij}+105\delta^6_{ij}) \\
	&\quad\,+\frac{35}{128}(\zeta^2_{ij} + \delta^2_{ij})^4\left\{\frac{\pi}{2} - \tan^{-1}\left(\frac{\delta_{ij}}{\zeta_{ij}}\right)\right\}
	 \end{aligned}
	\end{equation}
	When the resulting expression from combination of eqs. \ref{60}, \ref{61}, \ref{62} and \ref{63} is substituted in eq. \ref{57}, and integrated with respect to $\omega_{ij}$ from $0$ to $2\pi$, the terms containing $\zeta_{ij}\delta_{ij}$ and $\tan^{-1}(\delta_{ij}/\zeta_{ij})$ are eliminated. The final form of Eq. \ref{57} re-expanded in terms of $a_{ij}$ and $r_{ij}$ is given by
	\begin{equation} \label{64}
	\begin{aligned}
	w_i(r_{ij})+w_j(r_{ij})&=\bar{u}_{iz_{ij}}(r_{ij})+\bar{u}_{jz_{ij}}(r_{ij}) \\
	& = \left(\frac{1-\nu^2_i}{{E_i}}+\frac{1-\nu^2_j}{{E_j}}\right)\left(\frac{\pi}{4a_{ij}}\right)\left[a^2_{ij}\left(2\rho_1 + \frac{3}{2}\rho_2+\frac{5}{4}\rho_3+\frac{35}{32}\rho_4\right) - r^2_{ij}\left(\rho_1 + \frac{3}{2}\rho_2+\frac{15}{8}\rho_3\right.\right. \\
	&\quad\left.\left.+\frac{35}{16}\rho_4\right) + r^4_{ij}\left(\frac{9\rho_2}{16a^2_{ij}}+\frac{45\rho_3}{32a^2_{ij}}+\frac{315\rho_4}{128a^2_{ij}}\right)-r^6\left(\frac{25\rho_3}{64a^4_{ij}}+\frac{175\rho_4}{128a^4_{ij}}\right)+r^8_{ij}\left(\frac{1225\rho_4}{4096a^6_{ij}}\right)\right]
	\end{aligned}
	\end{equation}
	We now consider the three cases individually.
	
	\subsection{Case I: Two terms (N=2)} \label{secB.1}
	
	For $N=2$, terms with $\rho_3$ and $\rho_4$ are eliminated from eq. \ref{64}. Substituting the modified equation in eq. \ref{23} (Section \ref{sec4}) with two terms, we have
	\begin{equation} \label{65}
	\begin{aligned}
	&\left(\frac{1-\nu^2_i}{{E_i}}+\frac{1-\nu^2_j}{{E_j}}\right)\left(\frac{\pi}{4a_{ij}}\right)\left[a^2_{ij}\left(2\rho_1 + \frac{3}{2}\rho_2\right) - r^2_{ij}\left(\rho_1 + \frac{3}{2}\rho_2\right) + r^4_{ij}\left(\frac{9\rho_2}{16a^2_{ij}}\right)\right] \\
	&= (\gamma_{ij}+\gamma^{\mathrm{NL}}_{ij}) - \frac{r^2_{ij}\mathbb{A}_{ij}}{2}- \frac{r^4_{ij}\mathbb{B}_{ij}}{8}
	\end{aligned}
	\end{equation}
	In order to satisfy eq. \ref{65} for all points within the circular contact region, coefficients of like powers of $r_{ij}$ on both sides must be equal. Hence,
	\begin{equation} \label{66}
	-\left(\frac{1-\nu^2_i}{{E_i}}+\frac{1-\nu^2_j}{{E_j}}\right)\left(\frac{9\pi{\rho_2}}{8a^3_{ij}}\right) = \mathbb{B}_{ij}
	\end{equation}
	\begin{equation} \label{67}
	\left(\frac{1-\nu^2_i}{{E_i}}+\frac{1-\nu^2_j}{{E_j}}\right)\left(\frac{\pi}{2a_{ij}}\right)\left(\rho_1 + \frac{3}{2}\rho_2\right)=\mathbb{A}_{ij}
	\end{equation}
	\begin{equation} \label{68}
	\left(\frac{1-\nu^2_i}{{E_i}}+\frac{1-\nu^2_j}{{E_j}}\right)\left(\frac{\pi{a_{ij}}}{4}\right)\left(2\rho_1 + \frac{3}{2}\rho_2\right) = \gamma_{ij}+\gamma^{\mathrm{NL}}_{ij}
	\end{equation}
	Solving eqs. \ref{66} and \ref{67} for $\rho_1$ and $\rho_2$, we get
	\begin{equation} \label{69}
	\rho_1 = \frac{2a_{ij}(3\mathbb{A}_{ij}+2a^2_{ij}\mathbb{B}_{ij})}{3\pi}\left(\frac{1-\nu^2_i}{{E_i}}+\frac{1-\nu^2_j}{{E_j}}\right)^{-1}
	\end{equation}
	\begin{equation} \label{70}
	\rho_2 = -\frac{8a^3_{ij}\mathbb{B}_{ij}}{9\pi}\left(\frac{1-\nu^2_i}{{E_i}}+\frac{1-\nu^2_j}{{E_j}}\right)^{-1}
	\end{equation}
	Substituting the expressions for $\rho_1$ and $\rho_2$ obtained above into eq. \ref{68}, we get an expression for displacement $\gamma_{ij}$ in terms of contact radius $a_{ij}$
	\begin{equation} \label{71}
	\gamma_{ij}+\gamma^{\mathrm{NL}}_{ij} = a^2_{ij}\mathbb{A}_{ij} + \frac{a^4_{ij}}{3}\mathbb{B}_{ij}
	\end{equation}
	Also, by substituting the expressions for $\rho_1$ and $\rho_2$ in eq. \ref{53} and rearranging, we obtain a pressure distribution of the form	
	\begin{equation} \label{72}
	p_{ij}(r_{ij}) = \frac{2a_{ij}}{\pi}\left(\frac{1-\nu^2_i}{{E_i}}+\frac{1-\nu^2_j}{{E_j}}\right)^{-1}\left(1-\frac{r^2_{ij}}{a^2_{ij}}\right)^{1/2} \left[\mathbb{A}_{ij}+\frac{2a^2_{ij}\mathbb{B}_{ij}}{9}
	\left(1+2\frac{r^2_{ij}}{a^2_{ij}}\right)\right]
	\end{equation}
	
	\subsection{Case II: Three terms (N=3)} \label{secB.2}
	
	For $N=3$, terms with $\rho_4$ are eliminated. Substituting the modified eq. \ref{64} in eq. \ref{23} with three terms, we have
	\begin{equation} \label{73}
	\begin{aligned}
	&\left(\frac{1-\nu^2_i}{{E_i}}+\frac{1-\nu^2_j}{{E_j}}\right)\left(\frac{\pi}{4a_{ij}}\right)\left[a^2_{ij}\left(2\rho_1 + \frac{3}{2}\rho_2+\frac{5}{4}\rho_3\right) - r^2_{ij}\left(\rho_1 + \frac{3}{2}\rho_2+\frac{15}{8}\rho_3\right)+ r^4_{ij}\left(\frac{9\rho_2}{16a^2_{ij}}+\frac{45\rho_3}{32a^2_{ij}}\right)-r^6_{ij}\left(\frac{25\rho_3}{64a^4_{ij}}\right)\right] \\
	& = (\gamma_{ij}+\gamma ^{\mathrm{NL}}_{ij}) - \frac{r^2_{ij}\mathbb{A}_{ij}}{2}- \frac{r^4_{ij}\mathbb{B}_{ij}}{8}-\frac{r^6_{ij}\mathbb{C}_{ij}}{16}
	\end{aligned}
	\end{equation}
	Since the coefficients of like powers of $r_{ij}$ on both sides of eq. \ref{73} must be equal, we get
	\begin{equation} \label{74}
	\left(\frac{1-\nu^2_i}{{E_i}}+\frac{1-\nu^2_j}{{E_j}}\right)\left(\frac{25\pi{\rho_3}}{16a^5_{ij}}\right) = \mathbb{C}_{ij}
	\end{equation}
	\begin{equation} \label{75}
	-\left(\frac{1-\nu^2_i}{{E_i}}+\frac{1-\nu^2_j}{{E_j}}\right)\left(\frac{9\pi}{8a^3_{ij}}\right)\left(\rho_2+\frac{5{\rho_3}}{2}\right) = \mathbb{B}_{ij}
	\end{equation}
	\begin{equation} \label{76}
	\left(\frac{1-\nu^2_i}{{E_i}}+\frac{1-\nu^2_j}{{E_j}}\right)\left(\frac{\pi}{2a_{ij}}\right)\left(\rho_1 + \frac{3}{2}\rho_2+\frac{15}{8}\rho_3\right)=\mathbb{A}_{ij}
	\end{equation}
	\begin{equation} \label{77}
	\left(\frac{1-\nu^2_i}{{E_i}}+\frac{1-\nu^2_j}{{E_j}}\right)\left(\frac{\pi{a_{ij}}}{4}\right)\left(2\rho_1 + \frac{3}{2}\rho_2+\frac{5}{4}\rho_3\right) = \gamma_{ij}+\gamma^{\mathrm{NL}}_{ij}
	\end{equation}
	solving eqs. \ref{74}, \ref{75} and \ref{76} for $\rho_1$, $\rho_2$ and $\rho_3$, we get
	\begin{equation} \label{78}
	\rho_1 = \frac{2a_{ij}(15\mathbb{A}_{ij}+10a^2_{ij}\mathbb{B}_{ij}+9a^4_{ij}\mathbb{C}_{ij})}{15\pi}\left(\frac{1-\nu^2_i}{{E_i}}+\frac{1-\nu^2_j}{{E_j}}\right)^{-1}
	\end{equation}
	\begin{equation} \label{79}
	\rho_2 = -\frac{8a^3_{ij}(5\mathbb{B}_{ij}+9a^2_{ij}\mathbb{C}_{ij})}{45\pi}\left(\frac{1-\nu^2_i}{{E_i}}+\frac{1-\nu^2_j}{{E_j}}\right)^{-1}
	\end{equation}
	\begin{equation} \label{80}
	\rho_3 = \frac{16a^5_{ij}\mathbb{C}_{ij}}{25\pi}\left(\frac{1-\nu^2_i}{{E_i}}+\frac{1-\nu^2_j}{{E_j}}\right)^{-1}
	\end{equation}
	By substituting the expressions for $\rho_1$, $\rho_2$ and $\rho_3$ obtained above into eq. \ref{77}, we get
	\begin{equation} \label{81}
	\gamma_{ij}+\gamma^{\mathrm{NL}}_{ij} = a^2_{ij}\mathbb{A}_{ij}+\frac{a^4_{ij}}{3}\mathbb{B}_{ij}+\frac{a^6_{ij}}{5}\mathbb{C}_{ij}
	\end{equation}
	And by substituting the expressions for $\rho_1$, $\rho_2$ and $\rho_3$ into eq. \ref{53} and rearranging, we obtain a pressure distribution of the form
	\begin{equation} \label{82}
	p_{ij}(r_{ij}) = \frac{2a_{ij}}{\pi}\left(\frac{1-\nu^2_i}{{E_i}}+\frac{1-\nu^2_j}{{E_j}}\right)^{-1}\left(1-\frac{r^2_{ij}}{a^2_{ij}}\right)^{1/2}\left[\mathbb{A}_{ij}+\frac{2a^2_{ij}\mathbb{B}_{ij}}{9}\left(1+2\frac{r^2_{ij}}{a^2_{ij}}\right)+\frac{a^4_{ij}\mathbb{C}_{ij}}{25}\left(3+4\frac{r^2_{ij}}{a^2_{ij}}+8\frac{r^4_{ij}}{a^4_{ij}}\right)\right]
	\end{equation} 
	
	\subsection{Case III: Four terms (N=4)} \label{secB.3}
	
	For $N=4$, we substitute eq. \ref{64} into eq. \ref{23} to get
	\begin{equation} \label{83}
	\begin{aligned}
	&\left(\frac{1-\nu^2_i}{{E_i}}+\frac{1-\nu^2_j}{{E_j}}\right)\left(\frac{\pi}{4a_{ij}}\right)\left[a^2_{ij}\left(2\rho_1 + \frac{3}{2}\rho_2+\frac{5}{4}\rho_3+\frac{35}{32}\rho_4\right) - r^2_{ij}\left(\rho_1 + \frac{3}{2}\rho_2+\frac{15}{8}\rho_3+\frac{35}{16}\rho_4\right) \right.\\
	&\left.+r^4_{ij}\left(\frac{9\rho_2}{16a^2_{ij}}+\frac{45\rho_3}{32a^2_{ij}}+\frac{315\rho_4}{128a^2_{ij}}\right)-r^6_{ij}\left(\frac{25\rho_3}{64a^4_{ij}}+\frac{175\rho_4}{128a^4_{ij}}\right)+r^8_{ij}\left(\frac{1225\rho_4}{4096a^6_{ij}}\right)\right] \\ 
	&= (\gamma_{ij}+ \gamma^{\mathrm{NL}}_{ij}) - \frac{r^2_{ij}\mathbb{A}_{ij}}{2}- \frac{r^4_{ij}\mathbb{B}_{ij}}{8}-\frac{r^6_{ij}\mathbb{C}_{ij}}{16} - \frac{5r^8_{ij}\mathbb{D}_{ij}}{128}
	\end{aligned}
	\end{equation}
	Since the coefficients of like powers of $r_{ij}$ on both sides of eq. \ref{83} must be equal, we get
	\begin{equation} \label{84}
	-\left(\frac{1-\nu^2_i}{{E_i}}+\frac{1-\nu^2_j}{{E_j}}\right)\left(\frac{245\pi{\rho_4}}{128a^7_{ij}}\right) = \mathbb{D}_{ij}
	\end{equation}
	\begin{equation} \label{85}
	\left(\frac{1-\nu^2_i}{{E_i}}+\frac{1-\nu^2_j}{{E_j}}\right)\left(\frac{\pi}{16a^5_{ij}}\right)\left(25\rho_3+\frac{175}{2}\rho_4\right) = \mathbb{C}_{ij}
	\end{equation}
	\begin{equation} \label{86}
	-\left(\frac{1-\nu^2_i}{{E_i}}+\frac{1-\nu^2_j}{{E_j}}\right)\left(\frac{9\pi}{8a^3_{ij}}\right)\left(\rho_2+\frac{5{\rho_3}}{2}+\frac{35}{8}\rho_4\right) = \mathbb{B}_{ij}
	\end{equation}
	\begin{equation} \label{87}
	\left(\frac{1-\nu^2_i}{{E_i}}+\frac{1-\nu^2_j}{{E_j}}\right)\left(\frac{\pi}{2a_{ij}}\right)\left(\rho_1 + \frac{3}{2}\rho_2+\frac{15}{8}\rho_3+\frac{35}{16}\rho_4\right)=\mathbb{A}_{ij}
	\end{equation}
	\begin{equation} \label{88}
	\left(\frac{1-\nu^2_i}{{E_i}}+\frac{1-\nu^2_j}{{E_j}}\right)\left(\frac{\pi{a_{ij}}}{4}\right)\left(2\rho_1 + \frac{3}{2}\rho_2+\frac{5}{4}\rho_3+\frac{35}{32}\rho_4\right) = \gamma_{ij}+\gamma^{\mathrm{NL}}_{ij}
	\end{equation}
	Solving eqs. \ref{84}, \ref{85}, \ref{86} and \ref{87} for $\rho_1$, $\rho_2$, $\rho_3$ and $\rho_4$, we get
	\begin{equation} \label{89}
	\rho_1 = \frac{2a_{ij}(105\mathbb{A}_{ij}+70a^2_{ij}\mathbb{B}_{ij}+63a^4_{ij}\mathbb{C}_{ij}+60a^6_{ij}\mathbb{D}_{ij})}{105\pi}\left(\frac{1-\nu^2_i}{{E_i}}+\frac{1-\nu^2_j}{{E_j}}\right)^{-1}
	\end{equation}
	\begin{equation} \label{90}
	\rho_2 = -\frac{8a^3_{ij}(35\mathbb{B}_{ij}+63a^2_{ij}\mathbb{C}_{ij}+90a^4_{ij}\mathbb{D}_{ij})}{315\pi}\left(\frac{1-\nu^2_i}{{E_i}}+\frac{1-\nu^2_j}{{E_j}}\right)^{-1}
	\end{equation}
	\begin{equation} \label{91}
	\rho_3 = \frac{16a^5_{ij}(7\mathbb{C}_{ij}+20a^2_{ij}\mathbb{D}_{ij})}{175\pi}\left(\frac{1-\nu^2_i}{{E_i}}+\frac{1-\nu^2_j}{{E_j}}\right)^{-1}
	\end{equation}
	\begin{equation} \label{92}
	\rho_4 = -\frac{128a^7_{ij}\mathbb{D}_{ij}}{245\pi}\left(\frac{1-\nu^2_i}{{E_i}}+\frac{1-\nu^2_j}{{E_j}}\right)^{-1}
	\end{equation}
	By substituting the expressions for $\rho_1$, $\rho_2$, $\rho_3$ and $\rho_4$ obtained above into eq. \ref{88}, we get
	\begin{equation} \label{93}
	\gamma_{ij}+\gamma^{\mathrm{NL}}_{ij} = a^2_{ij}\mathbb{A}_{ij}+\frac{a^4_{ij}}{3}\mathbb{B}_{ij}+\frac{a^6_{ij}}{5}\mathbb{C}_{ij}+\frac{a^8_{ij}}{7}\mathbb{D}_{ij}
	\end{equation}
	And by substituting the expressions for $\rho_1$, $\rho_2$, $\rho_3$ and $\rho_4$ into eq. \ref{53} and rearranging, we obtain a pressure distribution of the form
	\begin{equation} \label{94}
	\begin{aligned}
	p_{ij}(r_{ij}) &= \frac{2a_{ij}}{\pi}\left(\frac{1-\nu^2_i}{{E_i}}+\frac{1-\nu^2_j}{{E_j}}\right)^{-1}\left(1-\frac{r^2_{ij}}{a^2_{ij}}\right)^{1/2} \left[\mathbb{A}_{ij}+\frac{2a^2_{ij}\mathbb{B}_{ij}}{9}
	\left(1+2\frac{r^2_{ij}}{a^2_{ij}}\right)+\frac{a^4_{ij}\mathbb{C}_{ij}}{25}\left(3+4\frac{r^2_{ij}}{a^2_{ij}}+8\frac{r^4_{ij}}{a^4_{ij}}\right)\right.\\
	&\quad\left.+\frac{4a^6_{ij}\mathbb{D}_{ij}}{245}\left(5+6\frac{r^2_{ij}}{a^2_{ij}}+8\frac{r^4_{ij}}{a^4_{ij}}+16\frac{r^6_{ij}}{a^6_{ij}}\right)\right]
	\end{aligned}
	\end{equation} 	
	
	\bibliographystyle{model5-names}
	\biboptions{authoryear}
	\bibliography{nlcurvcorrbibfile}
\end{document}